\documentclass[10pt,journal,compsoc]{IEEEtran}

\ifCLASSOPTIONcompsoc
  \usepackage[nocompress]{cite}
\else
  \usepackage{cite}
\fi
\ifCLASSINFOpdf
\else
\fi
\usepackage{amsmath}
\usepackage{algorithmic}
\usepackage[tight,footnotesize]{subfigure}
\usepackage{url}
\usepackage{amsthm}
\usepackage{amsfonts}

\usepackage{adjustbox}
\usepackage{algorithm}
\usepackage{color}
\usepackage{multirow}
\usepackage{setspace}
\usepackage{enumitem, mathtools}

\begin{document}
%
\title{FT-CNN: Algorithm-Based Fault Tolerance for Convolutional Neural Networks}
%
%
%
%

\author{Kai Zhao,
        Sheng Di,~\IEEEmembership{Senior,~IEEE,}
        Sihuan Li,
        Xin Liang,
        Yujia Zhai,
        Jieyang Chen,
        Kaiming Ouyang,
        Franck Cappello,~\IEEEmembership{Fellow,~IEEE},
        and Zizhong Chen,~\IEEEmembership{Senior,~IEEE}
\IEEEcompsocitemizethanks{
\IEEEcompsocthanksitem Kai Zhao, Sihuan Li, Yujia Zhai,and Zizhong Chen are with the Department of Computer Science and Engineering at University of California, Riverside, CA 92521.
\IEEEcompsocthanksitem Sheng Di and Franck Cappello are with the Mathematics and Computer Science division at Argonne National Laboratory, Lemont, IL 60439.
\IEEEcompsocthanksitem Xin Liang and Jieyang Chen are with the Computer Science and Mathematics Division at Oak Ridge National Laboratory, Oak Ridge, TN 37831.
}}

\IEEEtitleabstractindextext{%
\begin{abstract}
Convolutional neural networks (CNNs) are becoming more and more important for solving challenging and critical problems in many fields. CNN inference applications have been deployed in safety-critical systems, which may suffer from soft errors caused by high-energy particles, high temperature, or abnormal voltage. 
Of critical importance is ensuring the stability of the CNN inference process against soft errors. Traditional fault tolerance methods are not suitable for CNN inference because error-correcting code is unable to protect computational components, instruction duplication  techniques incur high overhead, and existing algorithm-based fault tolerance (ABFT) techniques cannot protect all convolution implementations. 
In this paper, we focus on how to protect the CNN inference process against soft errors as efficiently as possible, with the following three contributions. (1) We propose several systematic ABFT schemes based on checksum techniques and  analyze their fault protection ability and runtime thoroughly.
Unlike traditional ABFT based on matrix-matrix multiplication, our schemes support any convolution implementations. (2) We design a novel workflow integrating all the proposed schemes to obtain a high detection/correction ability with limited total runtime overhead. (3) We perform our evaluation using ImageNet with well-known CNN models including AlexNet, VGG-19, ResNet-18, and YOLOv2. Experimental results demonstrate that our implementation can handle soft errors with very limited runtime overhead 
(4\%$\sim$8\% in both error-free and error-injected situations).  
\end{abstract}

}

\maketitle

\IEEEdisplaynontitleabstractindextext

%
\IEEEpeerreviewmaketitle


%
%
%
%

\section{Introduction}
Deep learning using convolutional neural networks (CNNs) is becoming the key state-of-the-art technique in science and technology fields such as image classification\cite{alexnet, vgg-19, resnet50}, object detection\cite{yolov2}, natural language processing\cite{nlp}, medical image analysis\cite{medical_image}, and drug design\cite{drug}. More and more scientific research (such as cosmological simulation and materials analysis) also is addressing the great potential of leveraging CNN techniques to analyze extremely large amounts of data in a supercomputer environment, achieving unprecedented discoveries in their domains~\cite{wozniak2018candle}.

The reliability of the CNN inference is becoming a critical concern~\cite{overview_robust} because CNN inference applications are being widely utilized in different scenarios, including high-performance scientific simulations and safety-critical systems~\cite{critical1, critical2} such as aerospace and autonomous vehicles. CNN inference applications usually run for a long time or continuously to process many inference tasks. For example, the inference engine in autonomous vehicles is running continuously to predict road conditions. As a result, even a single inference task for one input finishes in seconds, the reliability of CNN inference is still critically important given the long execution time of the inference applications. 

In the domain of CNN inference, machine learning applications could be very error prone because of two reasons. 
On the one hand, recent literature indicates that soft errors are inevitable in modern  systems, from edge computing devices to supercomputers\cite{doe_sdc_report, largescale_sdc_simu}, because of multiple factors \cite{tan2015investigating} such as high-energy cosmic radiation~\cite{aerospaceabft}, aging, and wear of devices~\cite{geist2016kill}. On the other hand, CNN inference applications often call for power-efficient and cost-efficient machine learning accelerators, which~\cite{accelerator1, accelerator3} may adopt overclocking with voltage underscaling, incurring more soft errors than common hardware incurs.

Soft errors may cause serious consequences to CNN inference systems.
Recent studies \cite{Li_SC17, 8023727, ares} indicate that resilient convolutional neural networks are essential for guaranteeing the correctness of inference applications. 
Researchers \cite{Li_SC17} demonstrate that a single bit flip happened during CNN image classification could result in as much as 40\% and 70\% SDC rate in datapath and memory, respectively. Such high SDC rates would downgrade the CNN prediction accuracy dramatically.  
Furthermore, the neutron beam test \cite{8023727} shows that when running YOLO\cite{yolov2} classification, the Failure In Time (FIT) caused by SDCs could be as much as 38 for Nvidia K40 and 96 for Nvidia Tegra X1, which fail to meet the ISO 26262 standard for functional safety of road vehicles \cite{salay2017analysis}. 


Existing resilient solutions are insufficient for protecting  CNN inference applications against these soft errors. Error-correcting code (ECC), for example, suffers from memory area cost and relatively high latency and power consumption. According to \cite{ecc_overhead}, ECC with chip-kill applied to all data, compared with no ECC protection, has an average of 40\% overhead in memory energy, 20\% overhead in system energy and 20\% overhead in performance for computation-bounded applications. Moreover, ECC cannot handle multiple bit flips or computational errors. 
Techniques based on instruction duplication (ID)  \cite{vera2009selective} incur high overhead and require both application-specific and hardware-specific optimization; and optimizing and deploying  ID techniques on all CNN accelerators is difficult.


Considering all the drawbacks and limitations of ECC and ID, algorithm-based fault tolerance (ABFT) \cite{huang1984algorithm} is an attractive solution to realize resilient CNN. It has much lower overhead than other techniques have; and it is architecture independent, meaning that it supports any hardware accelerator. The idea of ABFT is to detect and/or correct soft errors based on the known invariants that the algorithm has. Over the past thirty years, ABFT schemes have been successful in  detecting errors for matrix operations \cite{wu2016,wu_hpdc16,tan_pc14, Chen:2018:FTO:3291656.3291747, chen2019tsm2, rivera2020tsm2x}, iterative methods \cite{Chen:2013:OOA:2442516.2442533, Tao:2016:NNO:2907294.2907306,tao2018improving,tao2018fault}, data transformation kernels\cite{Liang:2017:CSE:3126908.3126915} and sorting algorithm \cite{sihuan-ft-sorting}. 
However, the existing ABFT schemes for matrix operations focus mainly on large and square matrices. Moreover, they incur more than 50\% overhead when applied for CNN soft error protection (shown  by our experiments in Section \ref{sec:mmft_overhead} ).


In this paper, we propose a strategy comprising a series of ABFT schemes for protecting the CNN inference stage against soft errors. We focus on the convolutional layers in CNN because they consume the major portion of the computation time \cite{conv_time, eyeriss, jin2019deepsz, hu2020delta}.

The main contributions of this paper are summarized as follows.
\begin{itemize}
  \item We design several ABFT schemes that can be applied to any convolution implementation on any hardware. They can detect and correct errors at runtime. We provide an in-depth analysis of the ABFT schemes in terms of fault protection ability and runtime.
  \item We design a multischeme workflow for soft error protection with layerwise optimization to obtain a high detection/correction ability with limited runtime overhead. Additionally, our solution can protect the bias operation, grouped convolution, and back propagation.
  \item We implement an efficient soft error detection library for CNN, called \textit{FT-Caffe}, and evaluate FT-Caffe on ImageNet \cite{imagenet} using four popular CNN models: Alexnet \cite{alexnet}, VGG-19 \cite{vgg-19}, ResNet-18 \cite{resnet50}, and YOLOv2 \cite{yolov2}. Experimental results on the Bebop supercomputer \cite{bebop} using up to 128 nodes demonstrate that  FT-Caffe can keep the correctness of the inferences with 4\%$\sim$8\% overhead in both error-free and erroneous cases. 
\end{itemize} 

In the rest of the paper, we first introduce background about convolutional layers and existing ABFT techniques applicable to matrix-matrix multiplication (MM)-based convolution implementation. In Section~\ref{sec: novel schemes}, we propose four novel ABFT schemes that can be applied to any convolution implementations. In Section~\ref{sec: solutiton}, we analyze the fault protection ability and runtime of the four schemes and propose an efficient multischeme workflow integrating all the four schemes. In Section~\ref{sec: bias, grouped, back prop}, we discuss how to support bias, grouped convolution, and back propagation. In Section~\ref{sec: results}, we evaluate our solutions for both error-free case and erroneous case. In Section~\ref{sec: related work}, we discuss related work on fault tolerance in convolutional neural networks. We present our concluding remarks in Section ~\ref{sec: conclusion}.
\section{Background}
This section introduces some high-level ideas of convolutional layers and the existing ABFT techniques to MM-based convolution algorithms. The notations and symbols used in this paper are summarized in Table \ref{tab:parameters}.

\subsection{Definition of Convolutional Layer}
The convolutional layer can be represented as the following convolution operation.
\vspace{-1mm}
\begin{equation}
\begin{array}{l}
\hspace{-5mm}
\textbf{O}[n][m][x][y]=\textbf{B}[m]+\\
\hspace{-5mm}\sum\limits^{Ch-1}_{k=0}\sum\limits^{R-1}_{i=0}\sum\limits^{R-1}_{j=0}\textbf{D}[n][k][Ux+i][Uy+j] \times \textbf{W}[m][k][i][j]  \\
\hspace{-5mm}0\leq n< N,0\leq m<M,0\leq x,y<E,E=\frac{H-R+U}{U}
\end{array}
\hspace{-5mm}
\label{eq:conv} 
\end{equation}

The convolution operation involves two significant inputs: the feature map (fmap) $\textbf{D}$, $\textbf{D} \in \mathbb{R}^{N \times Ch\times H \times H}$, and the convolutional kernels $\textbf{W}$, $\textbf{W} \in \mathbb{R}^{M\times Ch \times R \times R}$. Note that all the matrices and vectors in this paper are highlighted in bold in order to differentiate from the scalar numbers, according to the naming convention. The bias, denoted as $\textbf{B}$, is applied to the output after convolution, and the final result is denoted as $\textbf{O}$, $\textbf{O} \in \mathbb{R}^{N\times M \times E \times E}$. Since the bias operation is independent of the convolution computation, in the rest of this section we describe only the protection for convolution computation. In Section \ref{subsec:bias_and_conv_group}, we will discuss the protection for bias.

\begin{table}[ht]
  \footnotesize
  \caption {Notations and Symbols Used in This Paper}
  \vspace{-4mm}
  \label{tab:parameters}
  \begin{adjustbox}{width=\columnwidth}
  \begin{tabular}{|c|c|}
  \hline
  Notation & Description \\ \hline
  \textbf{D} & Feature map, dimension is $4D$  \\ \hline
  \textbf{W} & Kernels, also called filters, dimension is $4D$  \\ \hline
  \textbf{O} & Output, dimension is $4D$   \\ \hline
  \textbf{B} & Bias, dimension is $1D$  \\ \hline
  \textbf{C} & Checksums \\ \hline
  \textbf{S} & Block summations of \textbf{O}, corresponding to checksums \\ \hline
  $\otimes$ & Convolution operation \\ \hline
  $N$ & First dimension of \textbf{D} and \textbf{O} \\ \hline
  $M$ & First dimension of \textbf{W} and second dimension of \textbf{O}\\ \hline
  $Ch$ & Second dimension of \textbf{D} and \textbf{W}, also called channels \\ \hline
  $H$ & Third and fourth dimension of \textbf{D} \\ \hline
  $R$ & Third and fourth dimension of \textbf{W} \\ \hline
  $E$ & Third and fourth dimension of \textbf{O}  \\ \hline
  $U$ & Stride size  \\ \hline
  \end{tabular}
  \end{adjustbox}
  \end{table}

\subsection{Implementation of Convolutional Layer}
\label{multi_conv_implementation}
 Convolution can be implemented efficiently in several ways~\cite{winograd}. The first option is \textbf{MM-based convolution}~\cite{chetlur2014cudnn}, which reshapes the kernel and feature map to two temporary matrices and then applies matrix-matrix multiplication (MM) on them.
 Another way to implement convolution is called \textbf{direct convolution}, which performs the convolution operation directly. It is widely used in AI accelerators including Eyeriss\cite{eyeriss}, DianNao\cite{diannao} and NVIDIA Deep Learning Accelerator\cite{nvdla}.
\textbf{Fast Fourier transform--based convolution} \cite{LIU2000101} leverages FFT to compute the convolution. It is particularly suitable for the relatively large feature map and kernel. However, it is inferior to the \textbf{Winograd convolution} \cite{winograd} when the sizes of the feature map and kernel are relatively small.




Modern CNN frameworks and accelerators generally automatically choose the best implementations of convolution based on hardware resources and model structure, because various implementations have different constraints on memory, architecture, and CNN model. 

\subsection{ABFT for Matrix-Matrix Multiplication}
\label{sec: abft_mm_intro}
Traditional ABFT designed for matrix-matrix multiplication can be applied to the MM calculation of the MM-based convolution \cite{8023727}, but it has at least three limitations. (1) It  supports only MM-based convolution implementation, which is not always the best-fit implementation selected by the CNN framework and accelerator. (2) It incurs high overhead (more than 50\%, as shown in Section \ref{sec:mmft_overhead}), due to the small and irregular shape of the matrices used by MM-based convolution. (3) Moreover, it cannot cover the reorganization operations of feature before the MM calculation. Therefore,  new ABFT schemes are needed in order to protect the convolutional layer more effectively. 




\section{Novel ABFT Schemes for Convolution}
\label{sec: novel schemes}

In this section, we present four novel ABFT schemes, each supporting any convolution implementation and being able to protect the whole convolution process. In Section \ref{sec: solutiton}, we propose a multischeme workflow using all the schemes in different stages to maximize the soft error protection ability with minimized performance overhead.

\subsection{Preliminary Analysis -- Convolution}
\begin{figure}[ht] \centering
{
\raisebox{0cm}{\includegraphics[scale=0.85]{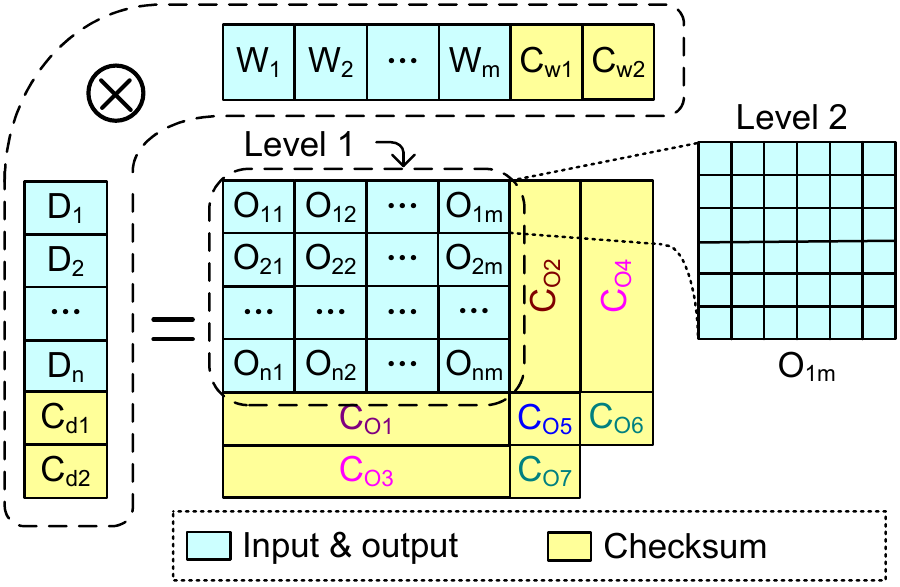}}
}
\vspace{-2mm}
\caption{Interpretation of Convolution at the Block Level}
\label{fig:convolution_as_basic_blocks}
\end{figure}

For clear description, we interpret convolution at the block level. 
Specifically, in Equation (\ref{eq:conv}), $\textbf{D}$, $\textbf{W}$, and $\textbf{O}$ are all $4D$ matrices. They can be represented as being composed of multiple blocks as shown in Figure \ref{fig:convolution_as_basic_blocks}.
For any $n$ and $m$ ($0\leq n<N,0\leq m<M$), $\textbf{D}_n$, $\textbf{W}_m$, and $\textbf{O}_{nm}$ are blocks. The dimension of $\textbf{D}_n$, $\textbf{W}_m$, and $\textbf{O}_{nm}$ are $Ch\times H \times H$, $Ch\times R \times R$ and $E\times E$, respectively.
The notation $\otimes$ is used to represent the convolution computation between blocks $\textbf{D}_n$ and $\textbf{W}_m$.
The convolution operation defined by Equation (\ref{eq:conv}) can be simplified at the block level as follows. 

\begin{equation}
\begin{array}{l}
\textbf{O}_{nm}=\textbf{D}_{n} \otimes \textbf{W}_m\\
0\leq n<N,0\leq m<M
\end{array}
\label{eq:conv_base}
\end{equation}

Equation (\ref{eq:conv_base}) can be interpreted by using the blue part in Figure \ref{fig:convolution_as_basic_blocks}. Since each of the 3D substructures of $\textbf{D}$ and $\textbf{W}$ is treated as a block, $\textbf{D}$ and $\textbf{W}$ can be thought of as two $1D$ vectors of blocks. At the block level, the convolution operation is similar to matrix-matrix multiplication. The element $(i,j)$ of $\textbf{O}$ is calculated by using the $i$th element of $\textbf{D}$ and the $j$th element of $\textbf{W}$. As illustrated in Figure \ref{fig:convolution_as_basic_blocks}, the elements involved in the convolutional layers can be split into two levels, which are covered by our protection solution, respectively.   

We can derive that the convolution operation (denoted by $\otimes$) has a distributive property as follows.
\begin{equation}
\label{eq:dis-prop}
\begin{array}{l}
\textbf{D}_1 \otimes \textbf{W}_1 + \textbf{D}_2 \otimes \textbf{W}_1= \\ 
\sum^{Ch-1}_{k=0}\sum^{R-1}_{i=0}\sum^{R-1}_{j=0}\textbf{D}_1[k][Ux+i][Uy+j]\times \textbf{W}_1[k][i][j]\\
+ \sum^{Ch-1}_{k=0}\sum^{R-1}_{i=0}\sum^{R-1}_{j=0}\textbf{D}_2[k][Ux+i][Uy+j]\times \textbf{W}_1[k][i][j]\\
\hspace{-1mm}=\hspace{-1mm} \sum^{Ch-1}_{k=0}\hspace{-1mm}\sum^{R-1}_{i=0}\hspace{-1mm}\sum^{R-1}_{j=0}(\textbf{D}_1\hspace{-1mm}+\hspace{-1mm}\textbf{D}_2)[k][Ux\hspace{-1mm}+\hspace{-1mm}i][Uy\hspace{-1mm}+\hspace{-1mm}j] \hspace{-1mm}\times \hspace{-1mm}\textbf{W}_1[k][i][j]\\\nonumber
\hspace{-1mm}=(\textbf{D}_1+\textbf{D}_2) \otimes \textbf{W}_1
\end{array}
\end{equation}
Similarly, we can get the following equation.
\begin{equation}
  \textbf{D}_1 \otimes \textbf{W}_1 + \textbf{D}_1 \otimes \textbf{W}_2 =\textbf{D}_1 \otimes (\textbf{W}_1 + \textbf{W}_2)
\end{equation}
The distributive property, Formula (\ref{eq:dis-prop}), is the key to proving the equivalence between the sum of the output and the output checksum. This property will be used later to prove the correctness of our design. 

\subsection{Preliminary Analysis -- CNN Checksums}

In general, we compute checksums for \textbf{D} and \textbf{W} and then use them to derive the checksums for \textbf{O}. Soft errors can be detected and corrected by comparing \textbf{O} with its checksums. 

We introduce all the checksums (as shown in Table \ref{tab:scheme_checksums} and as
yellow blocks in Figure \ref{fig:convolution_as_basic_blocks})  that are necessary for our ABFT schemes.
\begin{table}[ht]
\centering
\footnotesize
  \caption {Checksums Used by Schemes} 
  \vspace{-4mm}
  \label{tab:scheme_checksums} 
  \begin{adjustbox}{width=\columnwidth}
  \begin{tabular}{|c|c|c|c|}
  \hline
  \multirow{2}{*}{Scheme} & Checksums & Checksums \\
  &  of \textbf{D} and \textbf{W} & of \textbf{O} \\ \hline
  Full Checksum (FC) & $\textbf{C}_{d1}$, $\textbf{C}_{w1}$ & $\textbf{C}_{o1}$, $\textbf{C}_{o2}$ \\ \hline
  Row Checksum (RC) & $\textbf{C}_{d1}$, $\textbf{C}_{d2}$ & $\textbf{C}_{o1}$, $\textbf{C}_{o3}$ \\ \hline
  Column Checksum (ClC) & $\textbf{C}_{w1}$, $\textbf{C}_{w2}$ & $\textbf{C}_{o2}$, $\textbf{C}_{o4}$ \\ \hline
  Checksum-of-Checksum (CoC) & $\textbf{C}_{d1}$, $\textbf{C}_{w1}$, $\textbf{C}_{d2}$, $\textbf{C}_{w2}$ & $\textbf{C}_{o5}$, $\textbf{C}_{o6}$, $\textbf{C}_{o7}$ \\ \hline
  CoC Detection Only (CoC-D) & $\textbf{C}_{d1}$, $\textbf{C}_{w1}$, $\textbf{C}_{d2}$, $\textbf{C}_{w2}$ & $\textbf{C}_{o5}$ \\ \hline
\end{tabular}
\end{adjustbox}
\end{table}

We define the checksums of $\textbf{D}$ and $\textbf{W}$ as follows.
\begin{equation}
\begin{array}{l}
    \textbf{C}_{d1}=\sum^{N-1}_{n=0}\textbf{D}_n \\
    \textbf{C}_{d2}=\sum^{N-1}_{n=0}n \textbf{D}_n \\
    \textbf{C}_{w1}=\sum^{M-1}_{m=0}\textbf{W}_m \\
    \textbf{C}_{w2}=\sum^{M-1}_{m=0}m \textbf{W}_m \\
\end{array}
\end{equation}

The four checksums (denoted as input checksums) can be treated as four blocks of \textbf{D} and \textbf{W}. The checksums of \textbf{O} (denoted as output checksums) are defined as the convolution result of input checksums and/or inputs.
\begin{equation}
\begin{array}{l}
\textbf{C}_{o1} = \textbf{C}_{d1} \otimes \textbf{W} \\
\textbf{C}_{o2} = \textbf{D} \otimes \textbf{C}_{w1} \\
\textbf{C}_{o3} = \textbf{C}_{d2} \otimes \textbf{W}   \\
\textbf{C}_{o4} = \textbf{D} \otimes \textbf{C}_{w2} \\
\textbf{C}_{o5} = \textbf{C}_{d1} \otimes \textbf{C}_{w1} \\
\textbf{C}_{o6} = \textbf{C}_{d1} \otimes \textbf{C}_{w2} \\
\textbf{C}_{o7} = \textbf{C}_{d2} \otimes \textbf{C}_{w1} \\
\end{array}
\end{equation}

The output \textbf{O} is represented in the form of blocks (i.e., Level 1  in Figure \ref{fig:convolution_as_basic_blocks}). Elements inside the same block are independent with respect to checksums ( Level 2 in Figure \ref{fig:convolution_as_basic_blocks}). That is, we perform the checksum comparison independently for each element across blocks. Therefore, multiple soft errors in the same block can be detected and corrected independently.

In what follows, we describe the four schemes we proposed, each involving one or more input and output checksums. The required checksums used by each scheme are summarized in Table \ref{tab:scheme_checksums}. 

\subsection{Full Checksum  Scheme (FC)}
\label{sec full checksum}

The first scheme we designed is called \emph{full checksum scheme}, or FC, because it is based on checksums from both \textbf{D} and \textbf{W}, as shown in Figure \ref{fig:convolution_as_basic_blocks} and Table \ref{tab:scheme_checksums}.


$\textbf{C}_{d1}$ and $\textbf{C}_{w1}$ are calculated before the convolution operation, so any memory error striking $\textbf{D}$ or $\textbf{W}$ during the convolution would not affect $\textbf{C}_{d1}$ or $\textbf{C}_{w1}$. As for the output checksums, we can get the following equations by applying the distributive property of $\otimes$.
\begin{equation}
\begin{array}{l}
  \textbf{C}_{o1}[m]= \hspace{-1mm} (\sum\limits^{N-1}_{n=0} \textbf{D}_{n}) \otimes \textbf{W}_{m}=\hspace{-2mm} \sum\limits^{N-1}_{n=0} (\textbf{D}_{n}\otimes \textbf{W}_{m})=\hspace{-2mm} \sum\limits^{N-1}_{n=0} \textbf{O}_{nm}\\
  \textbf{C}_{o2}[n] \hspace{-1mm} = \hspace{-1mm} \textbf{D}_{n} \otimes (\sum\limits^{M-1}_{m=0} \textbf{W}_{m})=\hspace{-1mm} \sum\limits^{M-1}_{m=0} (\textbf{D}_{n}\otimes \textbf{W}_{m})=\hspace{-1mm} \sum\limits^{M-1}_{m=0} \textbf{O}_{nm} \nonumber
\end{array}
\end{equation}

These  equations show the equality between the sum of output and the output checksums. Let $\textbf{S}_{o1}$ and $\textbf{S}_{o2}$ be the summation of the output, where $\textbf{S}_{o1}[m]=\sum^{N-1}_{n=0} \textbf{O}_{nm}$ , $\textbf{S}_{o2}[n]=\sum^{M-1}_{m=0} \textbf{O}_{nm}$. We can compare $\textbf{C}_{o1}$, $\textbf{C}_{o2}$ with $\textbf{S}_{o1}$, $\textbf{S}_{o2}$ to detect, locate, and correct soft errors if they exist. 

\subsection{Row Checksum  Scheme (RC)}

Compared with the full checksum scheme, the second ABFT scheme we designed involves only the row checksums of output $\textbf{O}$, so we call it \emph{row checksum scheme}.

The row checksums used in this scheme are $\textbf{C}_{o1}$ and $\textbf{C}_{o3}$. $\textbf{C}_{o3}$ is computed from convolution operation between $\textbf{C}_{d2}$ and $\textbf{W}$, and the related output summation is defined by $\textbf{S}_{o3}[m]=\sum^{N-1}_{n=0} n\times \textbf{O}_{nm}$. 

For the detection of soft errors, we need to compare $\textbf{C}_{o1}$ with $\textbf{S}_{o1}$. If they are not equal to each other at location j, 
the error can be located by $i=\frac{\textbf{C}_{o3}[j]-\textbf{S}_{o3}[j]}{\textbf{C}_{o1}[j]-\textbf{S}_{o1}[j]}$ and $j$, and it can be corrected by adding $\textbf{C}_{o1}[j]-\textbf{S}_{o1}[j]$ to the block $(i,j)$.

\subsection{Column Checksum  Scheme (ClC) }
The third scheme we proposed is called \textit{column checksum scheme} because it involves only the column checksums of output $\textbf{O}$.
The column checksums used in this scheme are $\textbf{C}_{o2}$ and $\textbf{C}_{o4}$. $\textbf{C}_{o4}$ is defined by performing convolution operation between $\textbf{D}$ and $\textbf{C}_{w2}$,  and the related output summation is defined as $\textbf{S}_{o4}[n]=\sum^{M-1}_{m=0} m\times \textbf{O}_{nm}$. 
To detect soft errors, we compare $\textbf{C}_{o2}$ with $\textbf{S}_{o2}$ first. If they are not equal to each other at location $i$, the error can be located by $i$ and $j$ (=$\frac{\textbf{C}_{o4}[i]-\textbf{S}_{o4}[i]}{\textbf{C}_{o2}[i]-\textbf{S}_{o2}[i]}$), and it can be recovered by adding $\textbf{C}_{o2}[i]-\textbf{S}_{o2}[i]$ to the block $(i,j)$.

\subsection{Checksum-of-Checksum Scheme (CoC/CoC-D)}
\label{sec: coc-d}

Unlike the three schemes that all need $\textbf{D}$ and/or $\textbf{W}$ to calculate output checksums, the last scheme we proposed involves neither $\textbf{D}$ nor $\textbf{W}$ but only their checksums, so it is named  \textit{checksum-of-checksum scheme} (or CoC scheme for short). Specifically, $C_{o5}$, $C_{o6}$, and $C_{o7}$ are the output checksums we will use in this scheme.
Similar to $\textbf{C}_{o1}$, using the  distributive property can get three equations between the output checksums and output as follows.
\begin{equation}
  \textbf{C}_{o5} =\sum\nolimits_{n=0}^{N-1}\sum\nolimits_{m=0}^{M-1} \textbf{O}_{nm} =\textbf{S}_{o5}\nonumber
\end{equation}
\begin{equation}
  \textbf{C}_{o6} =\sum\nolimits_{n=0}^{N-1}\sum\nolimits_{m=0}^{M-1} m \times \textbf{O}_{nm} =\textbf{S}_{o6}\nonumber
\end{equation}  
\begin{equation}
  \textbf{C}_{o7} =\sum\nolimits_{n=0}^{N-1}\sum\nolimits_{m=0}^{M-1} n \times \textbf{O}_{nm} =\textbf{S}_{o7} \nonumber
\end{equation}

$S_{o5}, S_{o6}$, and $S_{o7}$ are defined as output summations corresponding to $C_{o5}, C_{o6},$ and $C_{o7}$. Let $\textbf{O}(i,j)$ be the corrupted output block, $\textbf{O}'$ be the correct output, and let $\delta = \textbf{O}'_{ij} - \textbf{O}_{ij}$ be the difference. Using the output checksums, we can get the following.

\begin{equation}
  \textbf{C}_{o5} -\textbf{S}_{o5} =\sum\nolimits_{n=0}^{N-1}\sum\nolimits_{m=0}^{M-1} \textbf{O}'_{nm}-\textbf{O}_{nm} = \delta\nonumber
\end{equation}
\begin{equation}
  \textbf{C}_{o6} -\textbf{S}_{o6} =\sum\nolimits_{n=0}^{N-1}\sum\nolimits_{m=0}^{M-1} m\times (\textbf{O}'_{nm}-\textbf{O}_{nm}) = j\times \delta \nonumber
\end{equation}
\begin{equation}
  \textbf{C}_{o7} -\textbf{S}_{o7} =\sum\nolimits_{n=0}^{N-1}\sum\nolimits_{m=0}^{M-1} n\times(\textbf{O}'_{nm}-\textbf{O}_{nm}) = i\times \delta\nonumber
\end{equation}

The location $i$, $j$ can be obtained by $i=(C_{o7} -S_{o7})/ \delta$ and $j=(C_{o6} -S_{o6}) / \delta$. Then the soft error can be fixed by adding $\delta$ to $\textbf{O}_{ij}$.

If only soft error detection is required, we do not need to compute $\textbf{C}_{o6}$ and $\textbf{C}_{o7}$, thus reducing the number of computations.
Input checksums     regarding $\textbf{C}_{d1}$, $\textbf{C}_{d2}$ and $\textbf{C}_{w1}$, $\textbf{C}_{w2}$, however,
are still required for soft error detection. We denote such a detection scheme by \emph{CoC-D}.
\section{Multischeme Workflow}
\label{sec: solutiton}
In this section, we first discuss the fault protection abilities and runtime of the four schemes we proposed in Section \ref{sec: novel schemes}. Then, we propose a multischeme workflow, powered by calibrated arrangement of the four schemes and layerwise optimization.

\subsection{Analysis of Protection Ability for Convolution Checksum Schemes}
In this section, we analyze the fault protection ability of all the schemes. 

\subsubsection{Fault Model}
The fault model for soft errors that we discuss in this paper includes transient faults in computational units and data corruption faults (both transient and persistent) in memory (including cache). In the following text, we use \emph{fault} to represent a malfunction event, and we denote its corresponding symptom  as \emph{soft error}. 

Soft error protection  includes error detection and error correction.  Error detection means that the scheme can detect soft errors without knowing the exact location. Error correction means that the scheme can locate the soft error locations and recover the incorrect result.

Without loss of generality, in the following analysis we consider at most one fault per convolution. One convolutional neural network contains several or even tens of convolutional layers, and the total forward execution time of a CNN model is usually within seconds. Thus, we can reasonably assume that at most one fault may strike to one convolutional layer, considering the short executing time of a single layer. Multiple faults per convolution can also be detected by our schemes and recovered by recomputing the corrupted convolutional layer.

\subsubsection{Analysis of Soft Error in \textbf{D} and \textbf{W}}

One fault occurring during the convolution execution can result in multiple soft errors in \textbf{W} and \textbf{D}. The soft errors in \textbf{W} can be detected by comparing the checksum of \textbf{W} with $\textbf{C}_{w1}$ and corrected by reloading weights from the CNN model. The soft errors in \textbf{D} do not need correction because \textbf{D} will be discarded after convolution computation; the resulting errors in the output can be detected and corrected by the checksums of the output, as demonstrated below.



\subsubsection{Analysis of Soft Error in \textbf{O}}
One fault during the convolution execution can result in corruption of one block row or column of \textbf{O}. 
By definition, the row $i$ of \textbf{O} is computed by the $i$th block of \textbf{D} with $W$. Thus, one fault in \textbf{D} would result in at most one corrupted row. The column $j$ of \textbf{O} is computed by \textbf{D} with the $j$th block of \textbf{W}. Thus, one fault in \textbf{W} would result in at most one corrupted column. Moreover, the intermediate result will be reused only by the same row or column, such that one fault in the computational units would  corrupt only values in the same row or column. Accordingly, in the following sections we discuss the soft error protection ability in the context of at most one corrupted row or column of \textbf{O}.

\subsubsection{Soft Error Protection Ability of CoC Scheme}

\begin{figure}[ht] \centering
\includegraphics[scale=0.63]{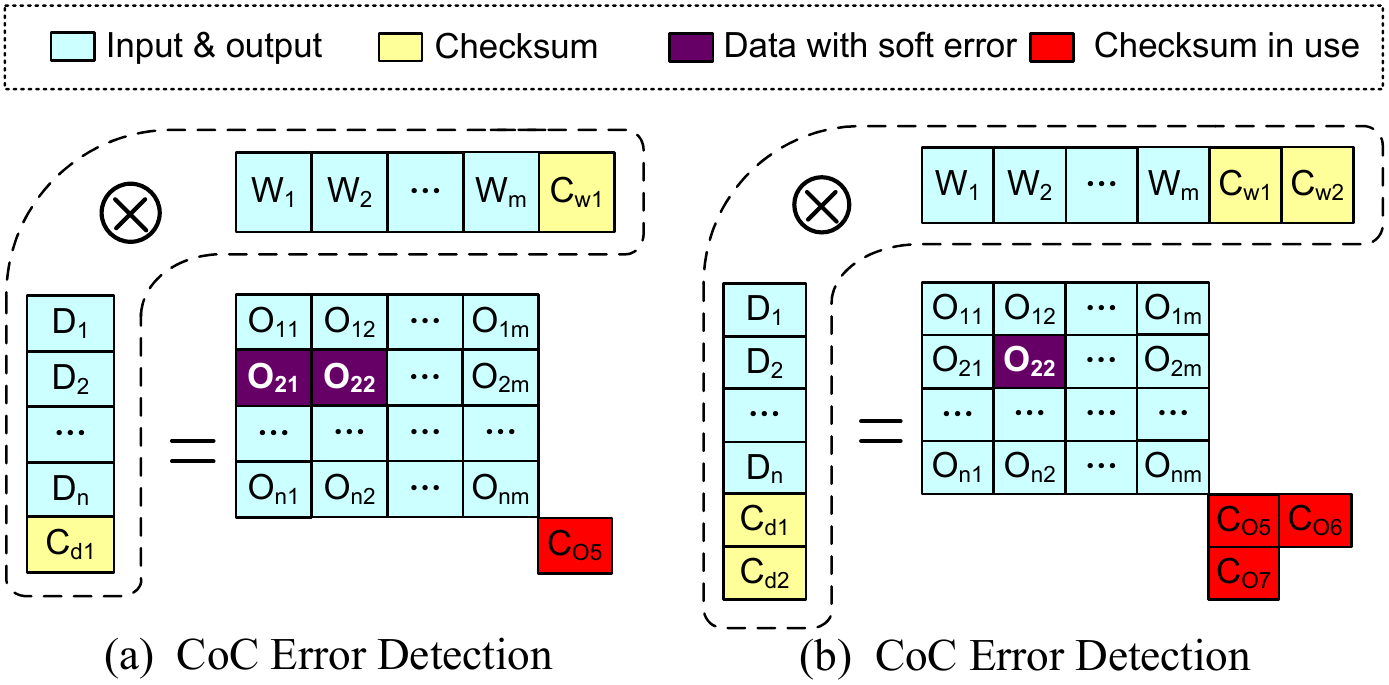}
\vspace{-2mm}
\caption{Soft Error Protection Ability of CoC Scheme (Soft error happens in inputs and outputs)}
\vspace{-2mm}
\label{fig:sdc_coc}
\end{figure}

Figure \ref{fig:sdc_coc} demonstrates the protection ability of the CoC scheme when soft errors strike the input or output data.
As shown in Figure \ref{fig:sdc_coc}(a), multiple soft errors can be detected by using only $\textbf{C}_{o5}$. A single soft error in $\textbf{O}$ can be corrected by CoC using all checksums including $\textbf{C}_{o5}$, $\textbf{C}_{o6}$, and $\textbf{C}_{o7}$, as shown in Figure \ref{fig:sdc_coc} (b). However, CoC cannot correct soft errors across multiple blocks in \textbf{O}. 


\begin{figure}[ht] \centering
\hspace{-8mm}
{
\raisebox{-1cm}{\includegraphics[scale=0.8]{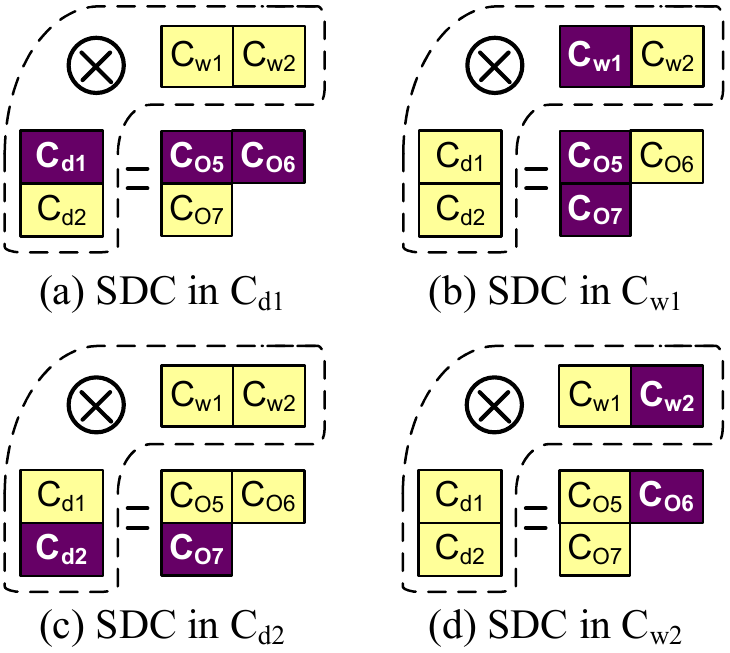}}
}
\hspace{-6mm}
\vspace{-2mm}
\caption{Soft Error Protection Ability of CoC Scheme (Soft error happens in checksums)}
\label{fig:sdc_inside_coc}
\end{figure}

Figure \ref{fig:sdc_inside_coc} illustrates the protection ability of the CoC scheme when soft errors happen inside the checksums. Such soft errors can cause inconsistency among the output checksums of CoC, which can be used for error detection. For example, in Figure \ref{fig:sdc_inside_coc}(a), $\textbf{C}_{d1}$ is corrupted, leading to corrupted $\textbf{C}_{o5}$ and $\textbf{C}_{o6}$ with correct $\textbf{C}_{o7}$. We can detect this abnormal pattern when comparing checksums with the summation of \textbf{O} to detect the input checksum corruption. The input \textbf{D}, \textbf{W}, and output \textbf{O} are clean and without soft errors since fault frequency is at most once per convolution. Thus, we can safely discard all the checksums and finish this convolution computation.

\subsubsection{Soft Error Protection Ability of Row Checksum Scheme and Column Checksum Scheme}

\begin{figure}[ht] \centering
\subfigure[{Row Checksum Scheme}]
{
\raisebox{-1cm}{\includegraphics[scale=0.65]{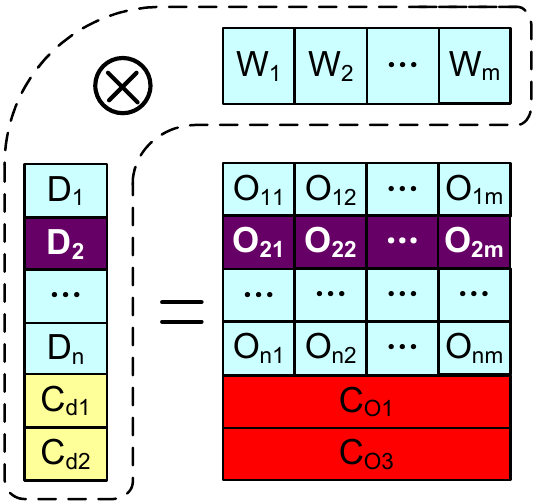}}
}
\subfigure[{Column Checksum Scheme}]
{
\raisebox{-1cm}{\includegraphics[scale=0.65]{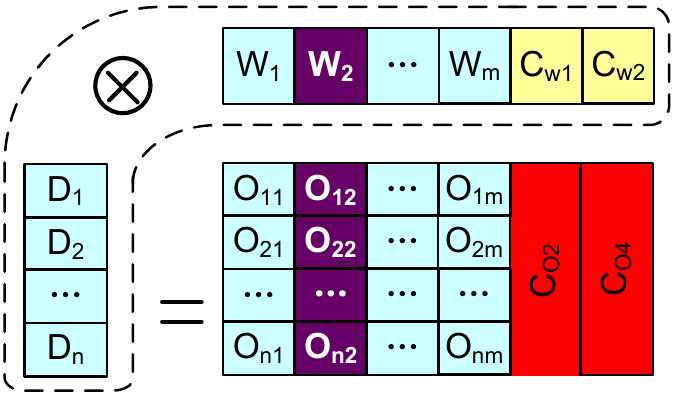}}
}
\vspace{-2mm}
\caption{Soft Error Protection Ability of Row/Column Checksum Schemes}
\label{fig:sdc_row_col}
\end{figure}

Since the row checksum scheme and column checksum scheme are symmetric with each other, we discuss them together in this section. As shown in Figure \ref{fig:sdc_row_col}(a), the row checksum scheme can detect and correct soft errors if they are in the same row. If the soft errors are in the same column, as shown in Figure \ref{fig:sdc_row_col}(b), the row checksum scheme can only detect soft errors; it has no correction ability. The column checksum scheme, on the contrary, can detect and correct errors located in the same column but fail to correct those appearing in the same row.

\subsubsection{Soft Error Protection Ability of Full Checksum Scheme}

\begin{figure}[ht] \centering
\subfigure[{Soft Error in the Same Row}]
{
\raisebox{-0.5cm}{\includegraphics[scale=0.7]{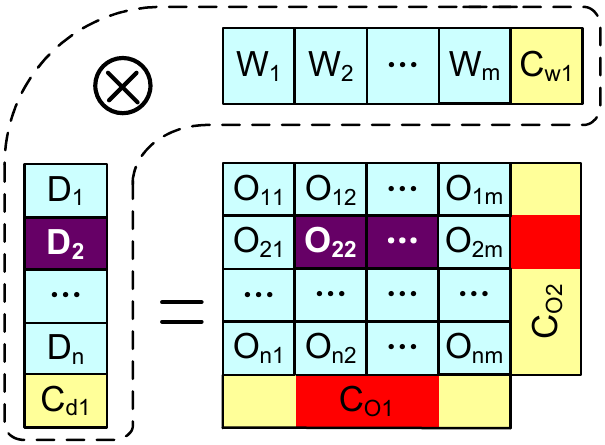}}
}
\hspace{-3mm}
\subfigure[{Soft Error in the Same Column}]
{
\raisebox{-0.5cm}{\includegraphics[scale=0.7]{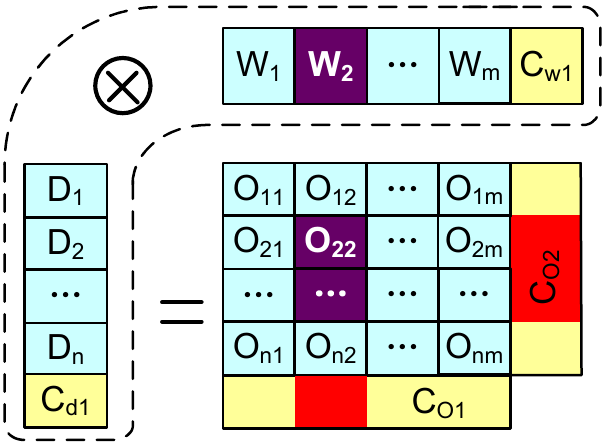}}
}

\hspace{-0mm}
\subfigure[{Soft error in the Same Row (including $\textbf{C}_{o2}$)}]
{
\raisebox{-1cm}{\includegraphics[scale=0.65]{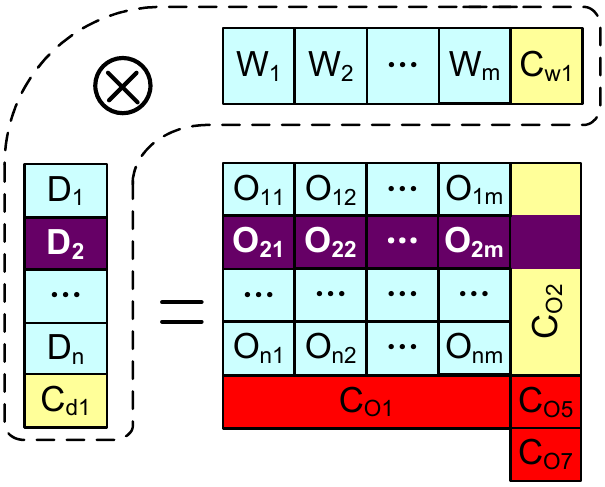}}
}
\hspace{-1mm}
\subfigure[{Soft error in the Same Column (including $\textbf{C}_{o1}$)}]
{
\raisebox{-1cm}{\includegraphics[scale=0.65]{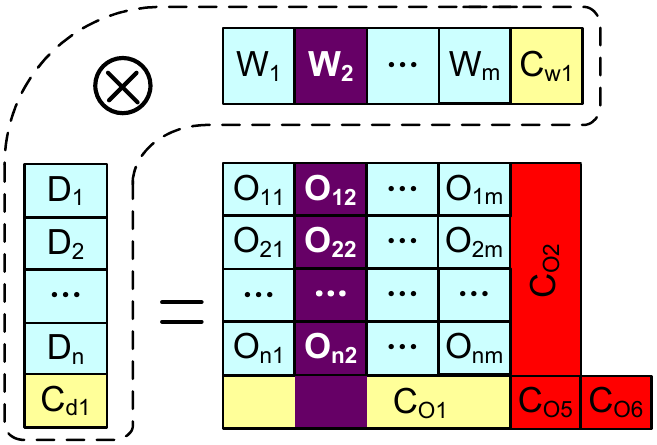}}
}
\vspace{-3mm}
\caption{Soft Error Protection Ability of Full Checksum Scheme}
\label{fig:sdc_full}
\end{figure}

The full checksum scheme has the highest ability to correct soft errors. The scheme uses both the row checksum $\textbf{C}_{o1}$ and column checksum $\textbf{C}_{o2}$ so that it can correct soft errors in both directions, as shown in Figure \ref{fig:sdc_full}(a)(b). 
If soft errors exist in $\textbf{C}_{o1}$ (Figure \ref{fig:sdc_full}(d)), however, $\textbf{C}_{o1}$ can no longer be used to locate or correct soft errors. To support error correction in this situation, we use checksum $\textbf{C}_{o5}$ and $\textbf{C}_{o6}$ from the CoC scheme to locate the corrupted column, and we then use $\textbf{C}_{o2}$ to correct the soft errors. 
If soft errors exist in $\textbf{C}_{o2}$ (Figure \ref{fig:sdc_full}(c)), $\textbf{C}_{o5}$ and $\textbf{C}_{o7}$ are used to locate the corrupted row, and $\textbf{C}_{o1}$ is used to correct the soft errors.

\subsubsection{Conclusion} In this section, we define our fault model and analyze the soft error protection ability of four schemes. We conclude that the CoC scheme has the lowest error correction ability and that the full checksum scheme has the best error correction ability. The abilities of the row checksum scheme and column checksum scheme are higher than that of the CoC scheme but lower than that of the full checksum scheme. CoC-D (discussed in Section \ref{sec: coc-d}) can detect multiple soft errors but without correction ability. The analysis here serves as the fundamental basis of our low-overhead high-protection design, which will be  presented in Section \ref{sec: solution-sub}.

\subsection{Runtime Analysis}
\label{sec: runtime-analysis}
In this section, we analyze the time complexity theoretically and present runtimes of all schemes based on experiments.

Table \ref{tab:conv_overhead} shows the time complexity of some basic checksum operations, where $\alpha$ is the coefficient of CPU-intensive operations and $\beta$ represents the coefficient for memory-intensive operations. 

\begin{table}[ht]
\footnotesize
\caption {Runtimes of Basic Operations} \label{tab:conv_overhead} 
\vspace{-3mm}
\begin{tabular}{|c|c|}
\hline 
\textbf{Operation} & \textbf{Derived Runtime}
\\ \hline
block level convolution $\textbf{D}_n \otimes \textbf{W}_m$ & $\alpha ChR^2E^2$    \\ \hline
Total convolution operations & $\alpha NMChR^2E^2$ \\ \hline
Compute the checksum of $\textbf{D}$  & $\beta NChH^2$ \\ \hline
Compute the checksum of $\textbf{O}$  & $\beta NME^2$ \\ \hline
\end{tabular}
\end{table}

Table \ref{tab:conv_overhead_comparison} shows the theoretical time complexity of all the schemes. The full checksum scheme has the best soft error correction ability; however, its runtime is relatively long. Although the CoC scheme has lower ability than the  other three schemes in correcting soft errors, it has the shortest runtime. Note that the kernel checksum $\textbf{C}_{w1}$ and $\textbf{C}_{w2}$ can be precalculated before the application; there is no cost in generating kernel checksum in the row, column, and CoC schemes.
\begin{figure}[ht] \centering
\hspace{-14mm}
\subfigure[{AlexNet}]
{
\raisebox{-1cm}{\includegraphics[scale=0.46]{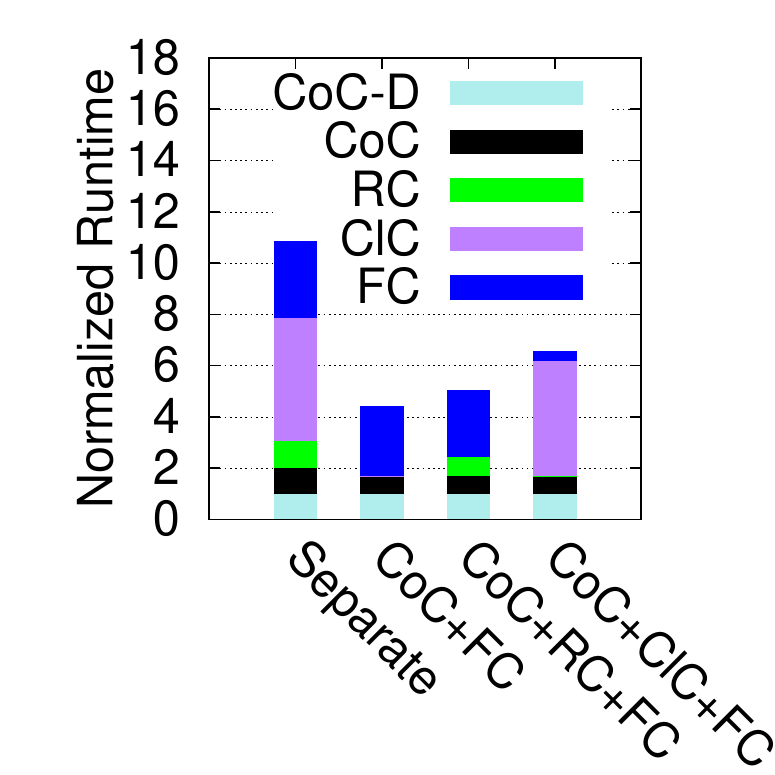}}
}
\hspace{-12mm}
\subfigure[{YOLOv2}]
{
\raisebox{-1cm}{\includegraphics[scale=0.46]{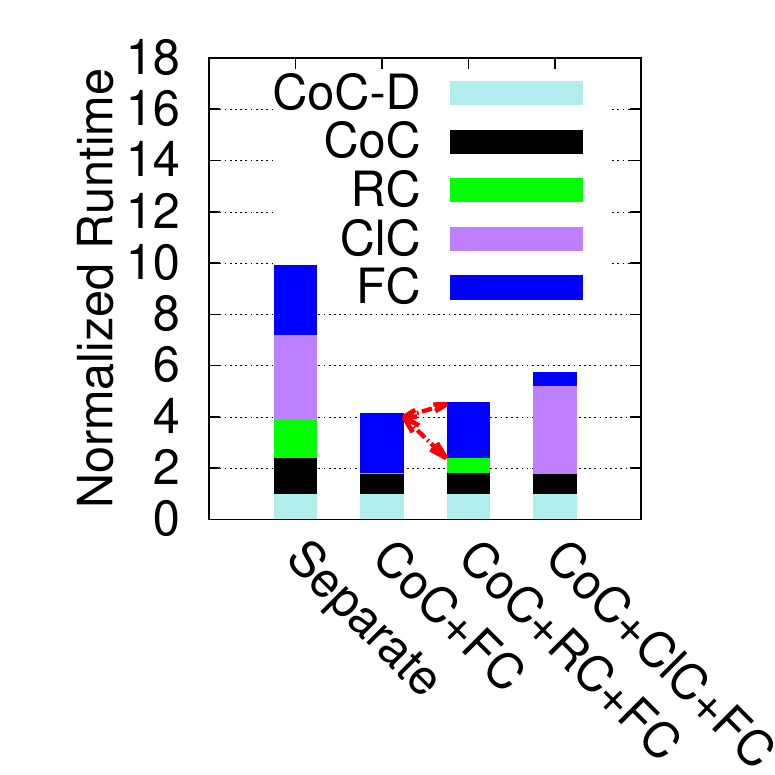}}
}
\hspace{-12mm}
\subfigure[{YOLOv2 (Conv8) }]
{
\raisebox{-1cm}{\includegraphics[scale=0.46]{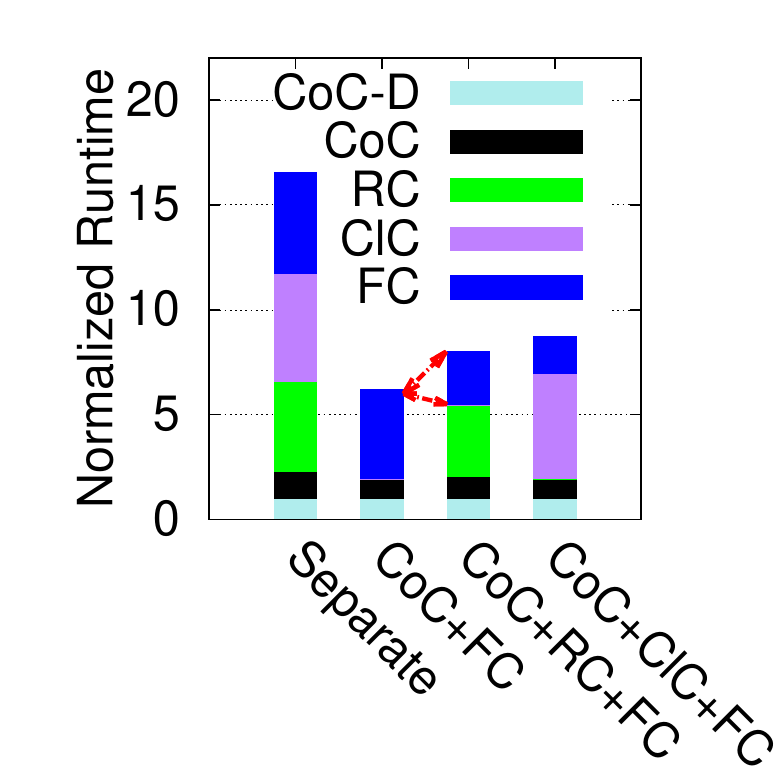}}
}
\hspace{-10mm}


\hspace{-14mm}
\subfigure[{VGG-19}]
{
\raisebox{-1cm}{\includegraphics[scale=0.46]{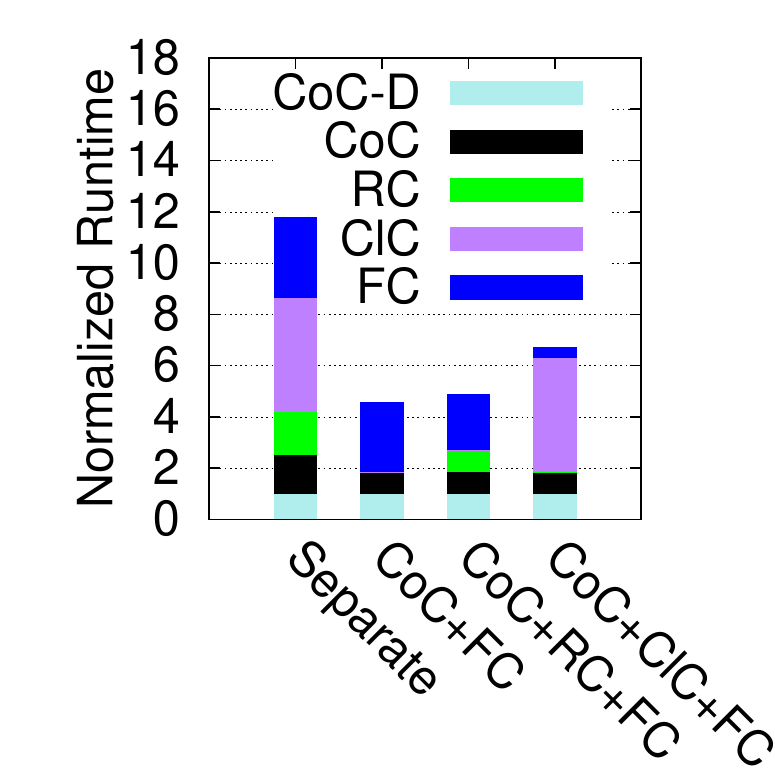}}
}
\hspace{-12mm}
\subfigure[{ResNet-18}]
{
\raisebox{-1cm}{\includegraphics[scale=0.46]{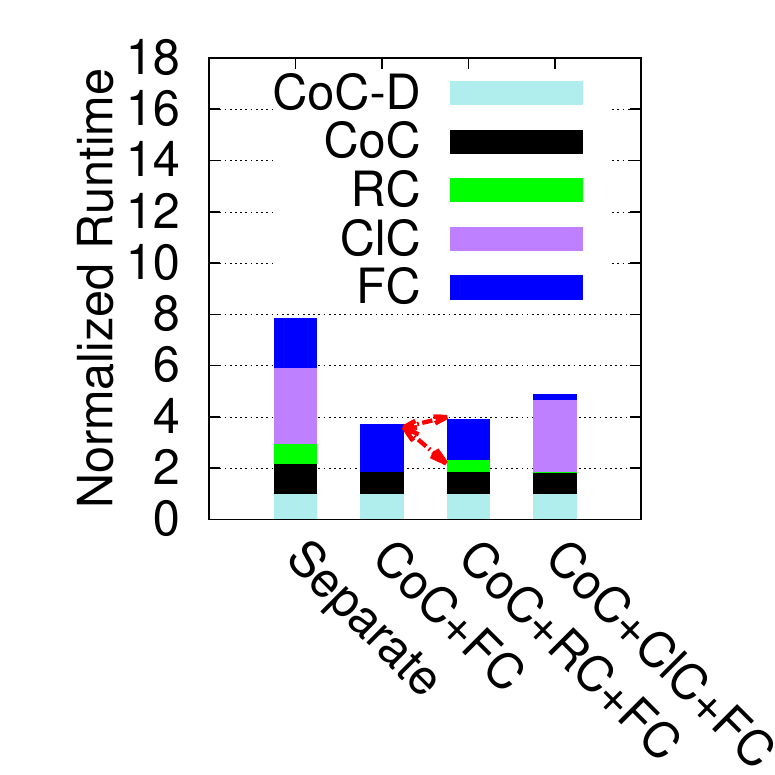}}
}
\hspace{-12mm}
\subfigure[{ResNet-18 (Conv1)}]
{
\raisebox{-1cm}{\includegraphics[scale=0.46]{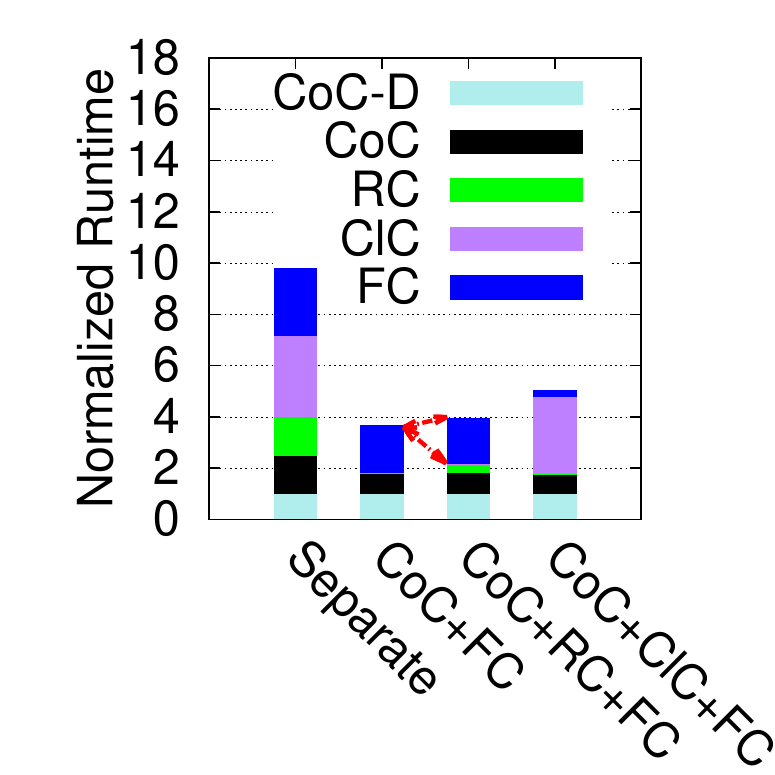}}
}
\hspace{-10mm}
\vspace{-3mm}
\caption{Worst-Case Normalized Runtime, Baseline is CoC-D}
\label{fig:anaysis-overhead}
\end{figure}

To verify the correctness of the derived time complexity of the four schemes, we execute them on a supercomputer using four CNN models. We show the normalized worst-case runtime of the four schemes in the $separate$ column of Figure \ref{fig:anaysis-overhead}. Other columns of this Figure represent the worst-case runtime of multischeme workflows and will be discussed in the next section.
Experiments confirm our conclusion that CoC and CoC-D have the shortest runtime and that the runtime of the full checksum scheme is relatively long. We also see that the column checksum scheme has a much longer runtime than the row checksum scheme does. The reason is twofold. On the one hand, $\textbf{W}$ blocks have smaller sizes than $\textbf{D}$ blocks have, leading to longer time to compute $\textbf{D}\otimes\textbf{C}_{w2}$  by the column checksum scheme  than to compute $\textbf{C}_{d2}\otimes \textbf{W}$  by the row checksum scheme. 
On the other hand, computing row checksums ($\textbf{C}_{o1}$ and $\textbf{C}_{o3}$) is more efficient than computing column checksums ($\textbf{C}_{o2}$ and $\textbf{C}_{o4}$), because the row checksum calculation can be reduced to efficient column-summation operations.

\begin{table}[ht]
  \centering
  \footnotesize
  \caption {ABFT Schemes Runtime} 
  \label{tab:conv_overhead_comparison} 
  \vspace{-3mm}
\begin{adjustbox}{width=\columnwidth}
  \begin{tabular}{|c|c|c|}
  \hline
  \multirow{2}{*}{\textbf{Scheme}} & \multirow{2}{*}{\textbf{Derived Runtime}} & \textbf{Soft Error} \\
  & & \textbf{Correction Ability} \\ \hline
  FC & $\alpha (N+M)ChR^2E^2 + \beta (NChH^2 + 2NME^2)$ & High \\ \hline
  RC & $ 2\alpha MChR^2E^2 + 2\beta (NChH^2+  NME^2) $ & Middle \\ \hline
  ClC & $ 2\alpha NChR^2E^2 + 2\beta ( NME^2) $ & Middle \\ \hline
  CoC & $ 3\alpha ChR^2E^2 + \beta (2NChH^2+3NME^2)$ & Low \\ \hline
\end{tabular}
\end{adjustbox}
\end{table}
\vspace{-2mm}

\subsection{Multischeme Workflow for Soft Error Protection}
\label{sec: solution-sub}
\begin{figure}[ht] \centering
\includegraphics[scale=0.55]{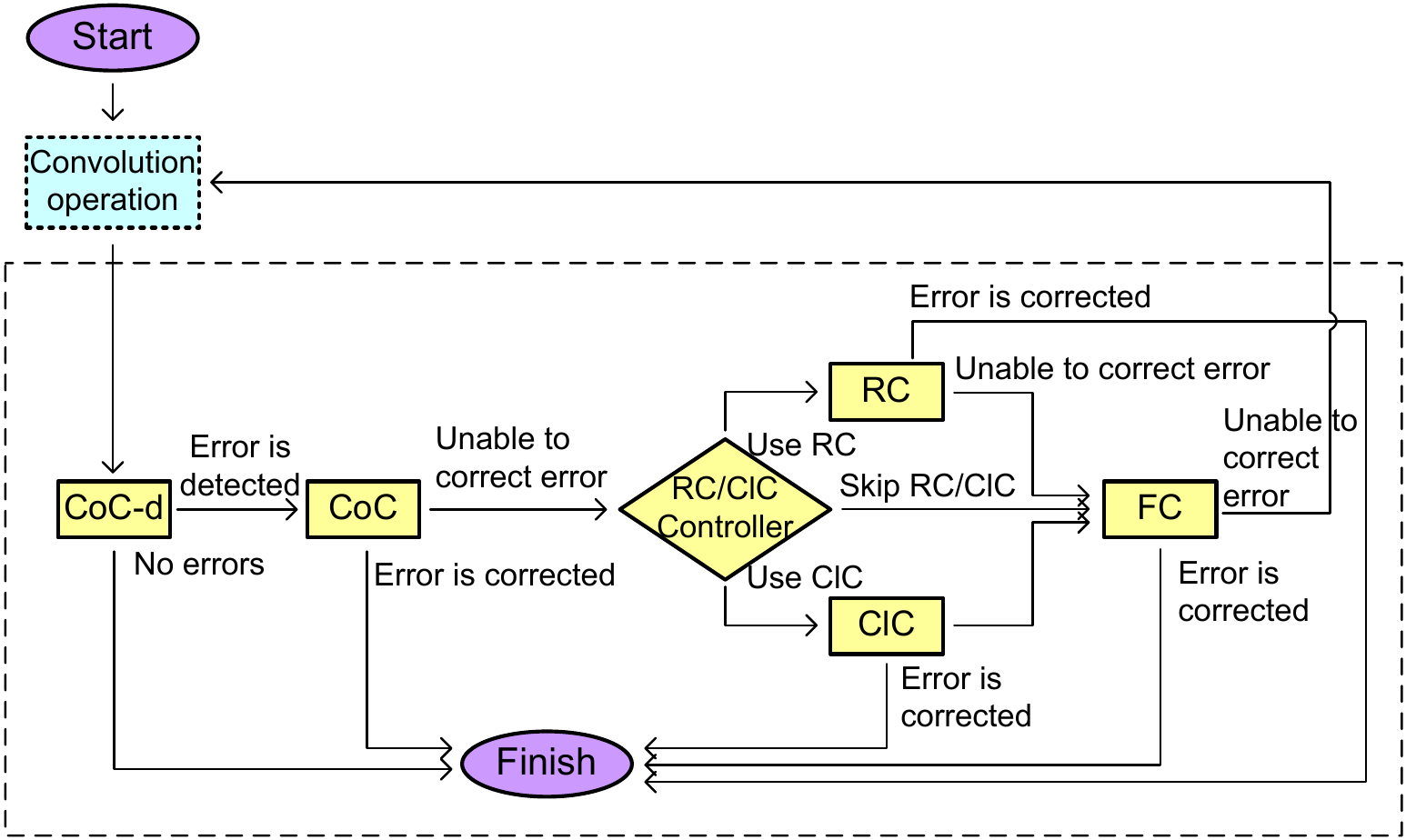}
\vspace{-2mm}
\caption{Multischeme Workflow Designed to Detect/Correct Soft Errors}
\label{fig:workflow}
\end{figure}

The four schemes we proposed have pros and cons in terms of their soft error correction ability and runtime overhead. To achieve the highest protection ability and lowest overhead, we propose a multischeme workflow by integrating the four schemes, as shown in Figure \ref{fig:workflow}. The workflow is made up of two modules: error detection and error correction. In our designed workflow, we use CoC-D to detect errors because it has the lowest overhead. For the error correction, we put CoC in the beginning because it is the most lightweight method. By comparison, FC has highest correction ability but also highest time overhead, so we put it at the end of the workflow.

The error detection modules will be executed for every execution whether there is a soft error or not. Thus, any unnecessary computations should be avoided in order to reduce the overall overhead. For instance, both CoC-D and FC are able to detect all the soft errors, but we adopt only CoC-D in the workflow for error detection because FC has a much higher overhead. RC and ClC cannot detect soft errors correctly if the checksum is corrupted.

The error correction module will not be executed until some soft errors are detected. The schemes in this module will be invoked to fix soft errors according to the workflow. If it fails to correct the errors due to inconsistency of checksum blocks or illegal error locations, the next-level scheme will be invoked. 

Since the checksums can be reused among different CNN schemes in the workflow, the runtime of the workflow is actually lower than the sum of all schemes' runtimes. For example, both CoC-D and CoC use $C_{o5}$; if CoC-D detects soft errors and CoC is invoked to correct soft errors, CoC can save the time of computing $C_{o5}$ and its corresponding summation $S_{o5}$, since they have been computed by CoC-D. This analysis can also be confirmed by our experiments. As shown in Figure \ref{fig:anaysis-overhead}, the relative runtime of CoC in the second column is reduced compared with that of CoC in the first column. The relative runtime of RC in the third column is reduced compared with that of RC in the first column.  

The decision to put RC and ClC in the workflow between CoC and FC is controlled by each layer.  
The reason to control RC/ClC in each layer is that their relative runtimes differ across layers.
Since RC and ClC are symmetric based, in the following we present our analysis based mainly on RC, without loss of generality.

We denote the runtime of the workflow CoC+FC as $t_0$, the runtime of workflow CoC+RC as $t_1$, and the runtime of workflow CoC+RC+FC as $t_2$. Enabling RC can fix some soft errors before FC, thus changing the runtime from $t_0$ to $t_1$.  When RC fails to correct soft errors, however, FC still needs to be invoked; and the runtime will increase from $t_0$ to $t_2$. Denoting the probability of row soft errors by $p_r$ and the probability of column soft errors by $p_c$, we can derive the average time saved by RC  as $t_y=p_r(t_0-t_1)$ and the average time increase by RC  as $t_n=p_c(t_2-t_0)$. In order to minimize the total runtime, RC should be enabled when $t_y > t_n$. 

We give an example to further illustrate when RC should be enabled. Figure \ref{fig:anaysis-overhead}(b) shows the average runtime among all the convolutional layers in YOLOv2. In this figure, the runtime of CoC+RC is much lower than that of CoC+FC, and the runtime of CoC+RC+FC is slightly higher than that of  CoC+FC. Therefore,  enabling RC  can save significant runtime when the soft errors are able to be corrected by RC. On the other hand,  a bit runtime penalty is incurred if RC fails to correct the soft errors. However, for the conv8 layer in YOLOv2 (shown in Figure \ref{fig:anaysis-overhead}(c)), CoC+RC's runtime is close to that of CoC+FC. Thus, enabling RC in this layer would barely reduce the overall runtime even though the soft errors can be corrected by RC. Moreover, CoC+RC+FC's runtime is much higher than CoC+RC's. As a result, the total runtime will increase significantly if the soft errors cannot be corrected by RC. Hence, for this layer, it is better to use CoC+FC for error correction with RC disabled. 

In practice, the runtime $t_0$, $t_1$ and $t_2$ can be computed by offline profiling. The probability values $p_c$ and $p_r$ can be estimated based on the size of \textbf{D} and size of \textbf{W}. For instance, the soft error often strikes each element in the input under the independent and identical distribution. In this situation, it is easy to drive that the probability of soft errors occurring in \textbf{D} is proportional to that of \textbf{W} (i.e., $\frac{p_r}{p_c}$ = $\frac{number\hspace{1mm} of \hspace{1mm}elements\hspace{1mm} in\hspace{1mm} \textbf{D}}{number \hspace{1mm}of\hspace{1mm} elements\hspace{1mm} in\hspace{1mm} \textbf{W}}$.

\section{\hspace{-4mm} Resolving Bias, Grouped Convolution, and Back Propagation}
\label{sec: bias, grouped, back prop}
In this section, we extend our solution to support bias, grouped convolution, and the back propagation of convolutional layers.

\subsection{Bias} \label{subsec:bias_and_conv_group}
Bias is a $1D$ vector that needs be added to the output of the convolutional layers.  FT-Caffe provides protection for the bias operation. 

Many CNN frameworks add bias on the fly with the convolution calculation. As a result, the output $\textbf{O}$ already contains bias, whereas the output checksums do not contain bias since they are calculated by inputs and input checksums without bias. In order to compare the output checksums and the output $\textbf{O}$, bias has to be subtracted from output summation before comparison. Subtracting bias from output $\textbf{O}$ directly before verification and then adding bias to $\textbf{O}$ after verification is not feasible, however, because of the overhead of modifying every element in $\textbf{O}$. Table \ref{tab:bias} shows the output checksums and adjusted output summation for comparison in order to detect errors. The bias part of the formulations can be precomputed.  

\begin{table}[ht]
\centering
\footnotesize
  \caption {Bias Adjustments for Output Checksums Comparison} 
  \vspace{-3mm}
  \label{tab:bias} 
  \begin{adjustbox}{width=0.8\columnwidth}
  \begin{tabular}{|c|c|}
  \hline
  Checksum & Adjust Summation \\ \hline
  $\textbf{C}_{o1}$ & $\textbf{S}_{o1}[m][i][j]- N\times Bias[m]$\\ \hline
  $\textbf{C}_{o3}$ & $\textbf{S}_{o3}[m][i][j]- (\sum^N_{i=1}i) \times Bias[m]$\\ \hline
  $\textbf{C}_{o2}$ & $\textbf{S}_{o2}[n][i][j]- \sum^m Bias[m]$\\ \hline
  $\textbf{C}_{o4}$ & $\textbf{S}_{o4}[n][i][j]- \sum^m m\times Bias[m]$\\ \hline
  $\textbf{C}_{o5}$ & $\textbf{S}_{o5}[i][j]- N\times\sum^m Bias[m]$\\ \hline
  $\textbf{C}_{o6}$ & $\textbf{S}_{o6}[i][j]- N\times\sum^m m\times Bias[m]$\\ \hline
  $\textbf{C}_{o7}$ & $\textbf{S}_{o7}[i][j]- (\sum^N_{i=1}i)\times\sum^m Bias[m]$\\ \hline

\end{tabular}
\end{adjustbox}
\end{table}

\subsection{ABFT for Grouped Convolution}
\begin{figure}[ht] \centering
\includegraphics[scale=.57]{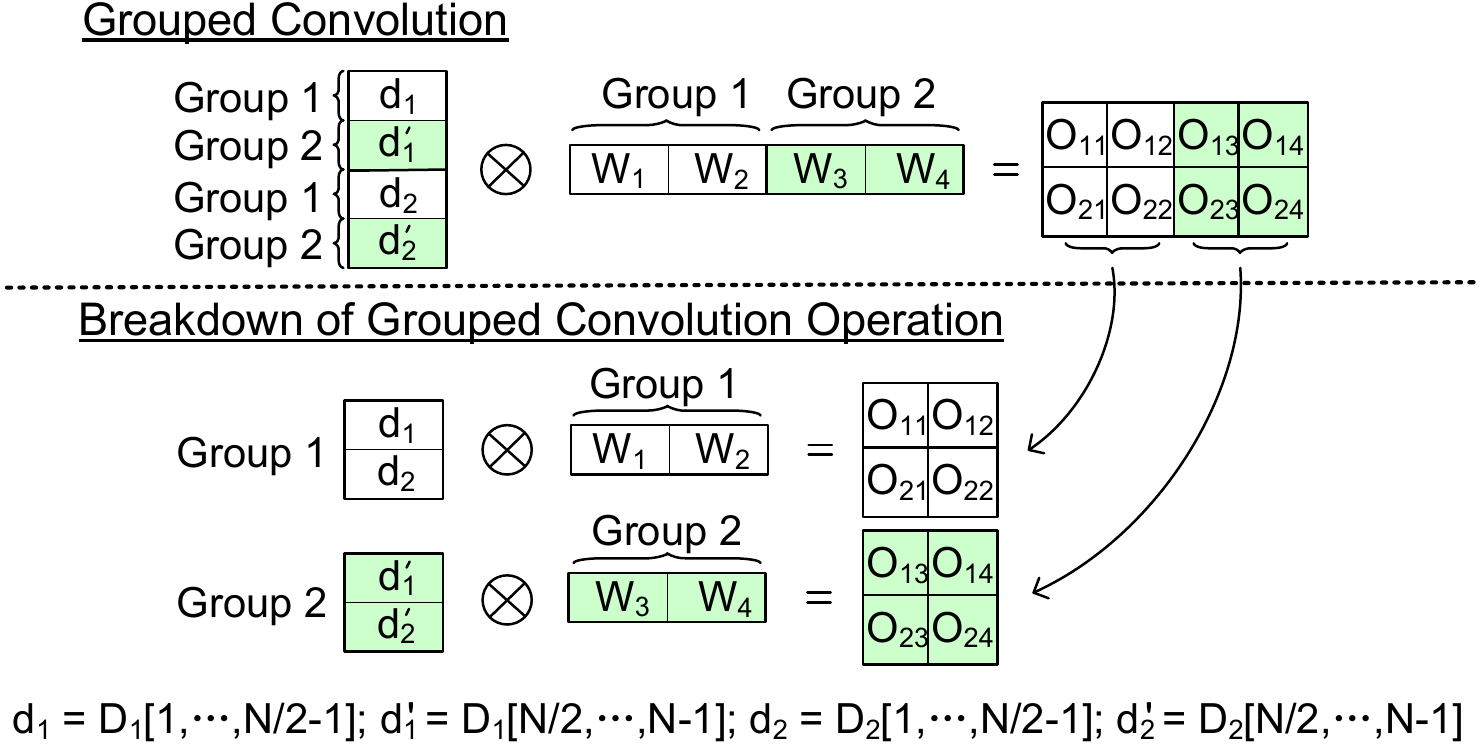}
\vspace{-4mm}
\caption{Demonstration of Grouped Convolution, Groups = 2}
\label{fig:grouped_conv}
\end{figure}

Grouped convolution is a special kind of convolution. Our schemes need to be modified to support this convolution.
Define the number of groups as $G$. Each fmap basic block has $\frac{Ch}{G}$ instead of $Ch$ channels. All the $M$ kernel basic blocks are divided into $G$ groups, each having $\frac{M}{G}$ $3D$ basic blocks. The kernel block in the $g$th group does convolution only with the $g$th channel group of every fmap block. Figure \ref{fig:grouped_conv} shows this process for N=2, M=4, and G=2.

The checksums for fmap $\textbf{C}_{d1}$ and $\textbf{C}_{d2}$ stay the same. The checksum for kernel are redefined as
\begin{equation}
  \textbf{C}_{w1}=[\sum^{\frac{M}{G}-1}_{m=0}\textbf{W}_m,\sum^{2\frac{M}{G}-1}_{m=\frac{M}{G}}\textbf{W}_m, ..., \sum^{M-1}_{m=(G-1)\frac{M}{G}}\textbf{W}_m ]\nonumber
\end{equation}
\begin{equation}
  \textbf{C}_{w2} =[\sum^{\frac{M}{G}-1}_{m=0}m\times \textbf{W}_m, \sum^{2\frac{M}{G}-1}_{m=\frac{M}{G}}m\times \textbf{W}_m, ...,\hspace{-5mm} \sum^{M-1}_{m=(G-1)\frac{M}{G}} \hspace{-3mm} m\times \textbf{W}_m ] \nonumber
\end{equation}
where $\textbf{C}_{w1}$ and $\textbf{C}_{w2}$ are the combination of $G$ checksums from each kernel group. Each checksum has $\frac{Ch}{G}$ channels, so $\textbf{Ch}_{w1}$ and $\textbf{Ch}_{w1}$ each have $G\frac{Ch}{G}=Ch$ channels, which are the same with every fmap block. 

The definition of output checksums $\textbf{C}_{o1}$, $\textbf{C}_{o2}$, ..., $C_{o7}$ stays the same. Let $X[l..r]$ represent the channels from $l$ to $r$ in matrix $X$.  We can prove the following property for any $\textbf{D}_n$ and $\textbf{W}_m$ according to Equation \eqref{eq:conv}.
\begin{equation}
\begin{array}{l}
  \hspace{-6mm} \textbf{D}_n \otimes \textbf{W}_m= \textbf{D}_n[1..k-1]\otimes \textbf{W}_m[1..k-1] \\
  \hspace{14mm}+ \textbf{D}_n[k..Ch] \otimes \textbf{W}_m[k..Ch], 0 \leq k < Ch\nonumber
\end{array}
\end{equation}
Using this equation, we can prove the relation between $\textbf{C}_{o2}$ and $\textbf{O}$  as follows. 
\begin{equation}
\begin{array}{l}
\textbf{C}_{o2}[n]=
\textbf{D}_n \otimes[\sum\limits^{\frac{M}{G}-1}_{m=0}\textbf{W}_m, \hspace{-2mm} \sum\limits^{2\frac{M}{G}-1}_{m=\frac{M}{G}}\hspace{-1mm}\textbf{W}_m, \cdots, \hspace{-3mm}\sum\limits^{M-1}_{m=(G-1)\frac{M}{G}}\hspace{-5mm}\textbf{w}_m ] \\
 = \textbf{D}_n[0..\frac{C}{G}\hspace{-0.5mm}-\hspace{-0.5mm}1] \hspace{-0.5mm}\otimes\hspace{-0.5mm} \hspace{-0.5mm}\sum\limits^{\frac{M}{G}-1}_{m=0}\hspace{-1mm}\textbf{w}_m
  \hspace{-0.5mm}+\hspace{-0.5mm} \textbf{D}_n[\frac{C}{G}..\frac{2C}{G}\hspace{-0.5mm}-\hspace{-0.5mm}1] \otimes \hspace{-2mm}\sum\limits^{\frac{2M}{G}-1}_{m=\frac{M}{G}}\hspace{-2mm}\textbf{w}_m\hspace{-0.5mm}\\
  + \cdots\nonumber 
  + \textbf{D}_n[\frac{(G-1)C}{G}..C-1] \otimes \sum\limits^{M-1}_{m=\frac{(G-1)M}{G}}\textbf{w}_m  \\
  = \sum\limits^{M-1}_{m=0} \textbf{D}_n \otimes \textbf{w}_m
  = \sum\limits^{M-1}_{m=0} \textbf{O}_{nm} \nonumber
\end{array}
\end{equation}

Similar equations can be proved for $\textbf{C}_{o1}$,$\textbf{C}_{o3}$,$\textbf{C}_{o4}$, $C_{o5}$, $C_{o6}$, and $C_{o7}$. Therefore, all the ABFT schemes we proposed can be applied to grouped convolution.  

\subsection{ABFT for Convolution Back Propagation}
Our schemes can also be applied to back propagation together with forward pass so that the convolutional layers can be fully protected in the training phase. 

\begin{figure}[ht] \centering
\includegraphics[scale=.5]{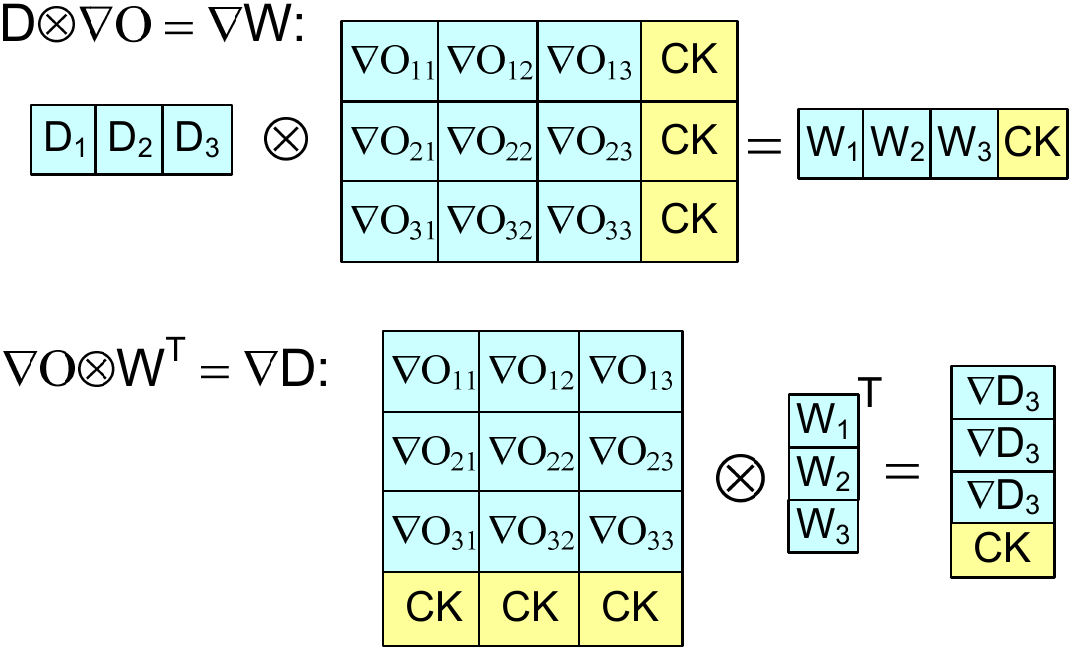}
\vspace{-4mm}
\caption{Demonstration of Checksum Design for Back Propagation}
\label{fig:abft_for_back_propagation}
\end{figure}

During back propagation, the gradient of kernel $\nabla \textbf{W}$ is used by methods such as gradient descent in order to update $\textbf{W}$. The gradient of fmap $\nabla \textbf{D}$ is used to get $\nabla \textbf{O}$ of the previous layer. As shown in Figure \ref{fig:abft_for_back_propagation}, the gradients are calculated as $\textbf{D} \otimes \nabla \textbf{O}= \nabla \textbf{W}$ and $\textbf{W}^T \otimes \nabla \textbf{O}= \nabla \textbf{D}$. Checksums for $\nabla \textbf{O}$ are used in this situation to protect the two convolution operations.

Since CNN models are usually trained in a more stable environment than the inference stage and since the training stage can tolerate some soft errors because of their iterative-convergent nature, we focus our experiments on the inference stage.

\section{Experimental Evaluation}
In this section, we evaluate our multischeme workflow using our FT-Caffe fault tolerance CNN framework.

\label{sec: results}

\subsection{Experimental Setup}

\emph{FT-Caffe.} Our FT-Caffe framework is based on Intel-Caffe. MKL-DNN is enabled to support dynamic selection of convolution execution. MKL-DNN contains all the convolution implementations we discussed in Section \ref{multi_conv_implementation}. It automatically chooses the most suitable implementation to use for each convolutional layer. To compare the runtime overhead of our solution with that of the ABFT designed for matrix-matrix multiplication, we also perform the experiments based on the  MM-based convolution implementation.

\emph{CNN models and dataset.} We tested our FT-Caffe with four widely used networks: AlexNet, VGG-19, ResNet-18, and YOLOv2. Pretrained Caffe models are used together with model prototypes for deployment. We adopt the ImageNet validation set, which contains 50k images. The images are preprocessed to smaller size in order to save picture processing time when the program starts. The batch size is set to 64. 

\emph{Experimental platforms.} We conducted our experiments on the Bebop supercomputer\cite{bebop} at Argonne National Laboratory using up to 128 nodes. Each node is equipped with 128 GB memory and two Intel Xeon E5-2695 v4 processors (each with 16 cores)

\emph{Error injection.} 
To demonstrate the overhead of our fault tolerant solutions, we inject soft errors at the source code level as most ABFT works did \cite{Liang:2017:CSE:3126908.3126915, Chen:2018:FTO:3291656.3291747}. The consequences of one computational fault or memory fault are simulated by randomly corrupting selected row or column of output. We denote the total number of convolutional layers of a CNN model as $L$. To assess the overhead accurately, we run the experiments for $L$ epochs corresponding to the numbers of convolutional layers of each network ($L$= 5, 9, 16, 21 for AlexNet, YOLOv2, VGG-19, and ResNet-18, respectively). For the $i$th epoch, we inject errors to $i$th convolutional layer. The final overhead is the arithmetic mean of all the inference executions and the standard deviation in our experiments is within 5\%. 

\subsection{Experimental Results with MKL-DNN}
\label{sec:exp_result}

In this section, we present our evaluation results with MKL-DNN . We analyze the results from the perspective of execution time overhead for both error-free cases and erroneous cases.

\begin{figure}[ht] \centering
\hspace{-9mm}
\subfigure[{Error-free}]
{
\raisebox{-1cm}{\includegraphics[scale=0.4]{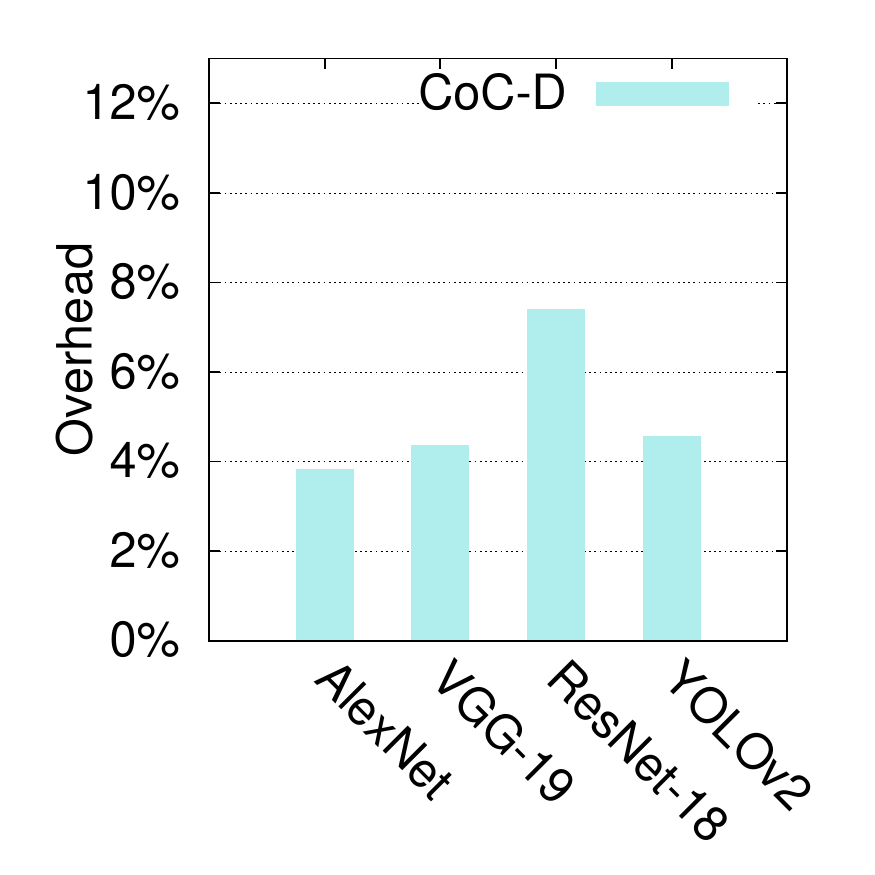}}
}
\hspace{-9mm}
\subfigure[{\hspace{9mm}Erroneous \newline \hspace*{8mm}(RC/ClC Disabled)}]
{
\raisebox{-1cm}{\includegraphics[scale=0.4]{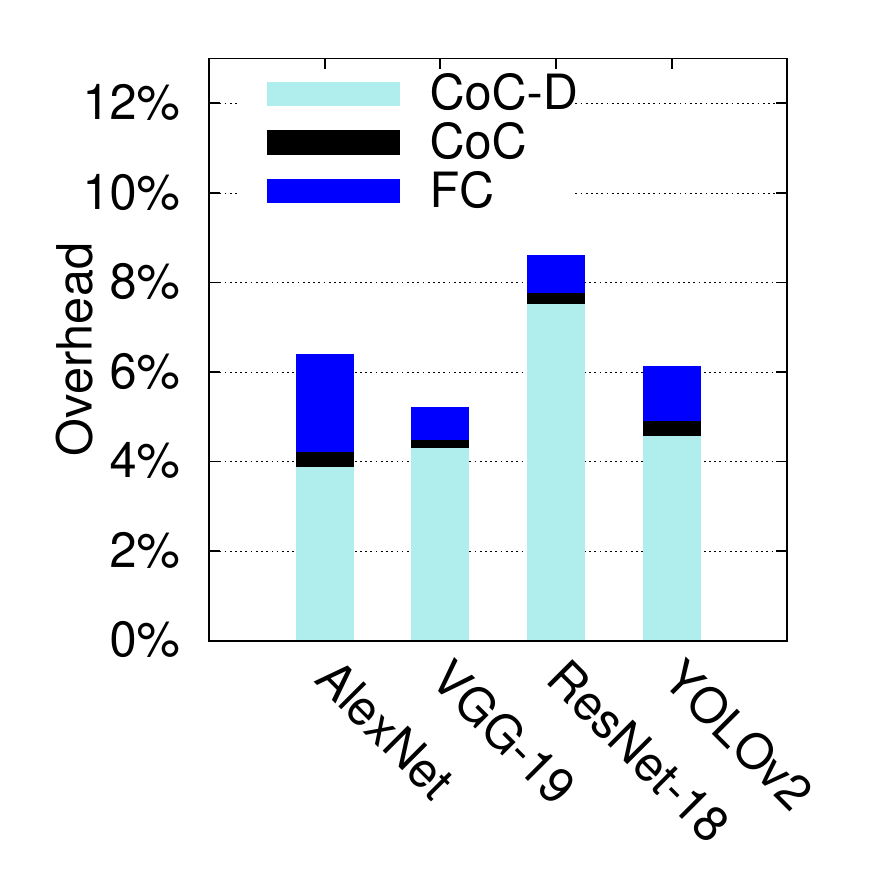}}
}
\hspace{-9mm}
\subfigure[{\hspace{9mm}Erroneous \newline \hspace*{8mm} (RC/ClC enabled)}]
{
\raisebox{-1cm}{\includegraphics[scale=0.4]{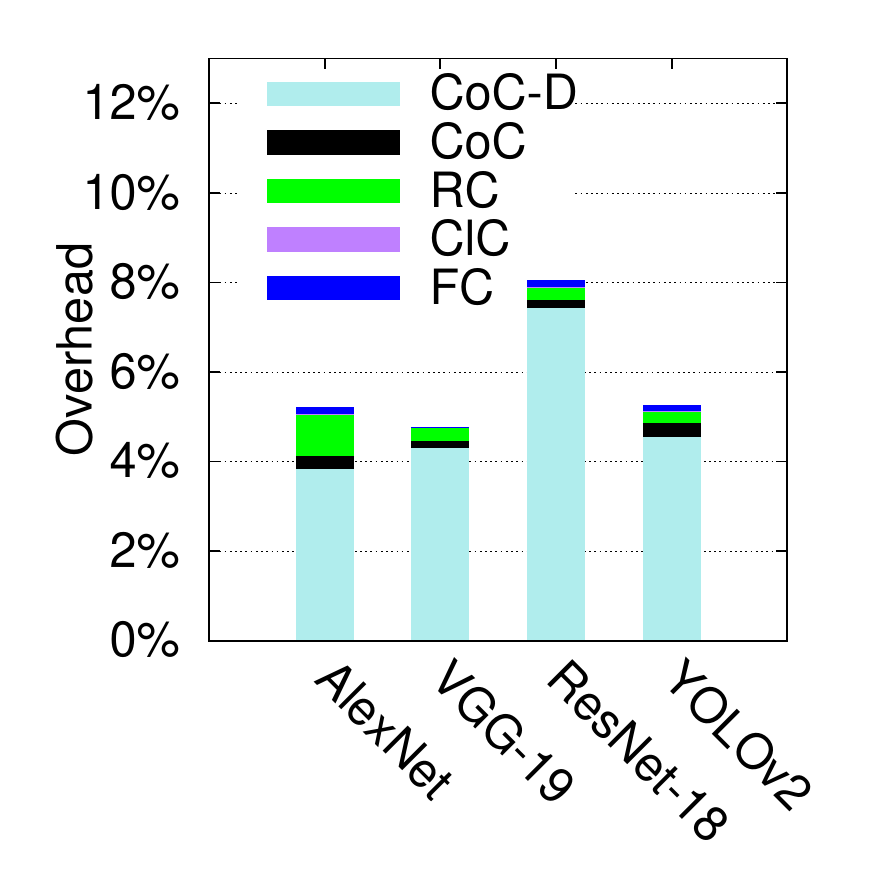}}
}
\hspace{-8mm}
\vspace{-4mm}
\caption{Runtime Overhead with MKL-DNN}
\label{fig:overhead_mkldnn}
\end{figure}

\emph{Error-free cases.} The experimental results in the error-free cases are presented in Figure \ref{fig:overhead_mkldnn}(a). 
We can see from the figure that our FT-caffe can protect the inference stage with less than 4\%, 4.5\%, 8\%, and 5\% overhead for AlexNet, VGG-19, YOLOv2, and ResNet-18, respectively, regardless of the convolution implementation. These results show that our CoC-D error detection scheme has relatively short runtime compared  with the convolution execution, which is attributed to the design of avoiding unnecessary computations in our solution (see Section \ref{sec: solution-sub}). The reason ResNet-18 has higher overhead than the other models have is that the ResNet-18 has small convolution kernels (\textbf{W} size is $M$$\times$$C$$\times$$ 3$$\times$$3$) in all the convolutional layers, which have relatively short computing time; thus, the checksum computation and verification time percentage would be relatively large. 

\emph{Erroneous cases -- RC/ClC disabled.} To show the effectiveness of our layerwise optimization for RC/ClC, we first test our multischeme workflow with RC/ClC disabled in erroneous cases. Figure \ref{fig:overhead_mkldnn}(b) demonstrates that the runtime overheads (including both error detection and error correction) of the four CNN models are all below 9\%. The error detection overhead is higher than the error correction overhead because the error detection scheme is executed for every convolution operation whereas the error correction schemes are invoked only when errors are detected. The full checksum scheme dominates the error correction overhead, thus confirming our analysis in Section \ref{sec: solutiton} that FC has high protection ability and relatively longer runtime.

\begin{figure}[ht] \centering
\vspace{-6mm}
\hspace{-4mm}
\subfigure[Distribution of  different workflows]
{
\raisebox{-1cm}{\includegraphics[scale=0.45]{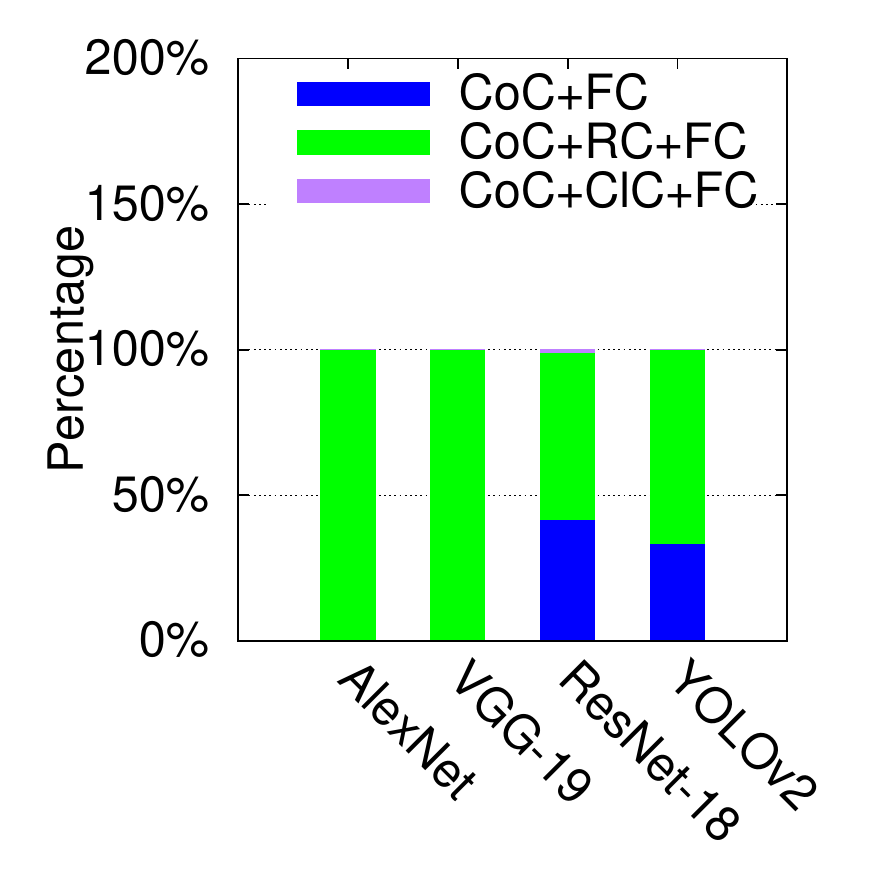}}
}
\hspace{4mm}
\subfigure[{Distribution of soft errors corrected by schemes}]
{
\raisebox{-1cm}{\includegraphics[scale=0.45]{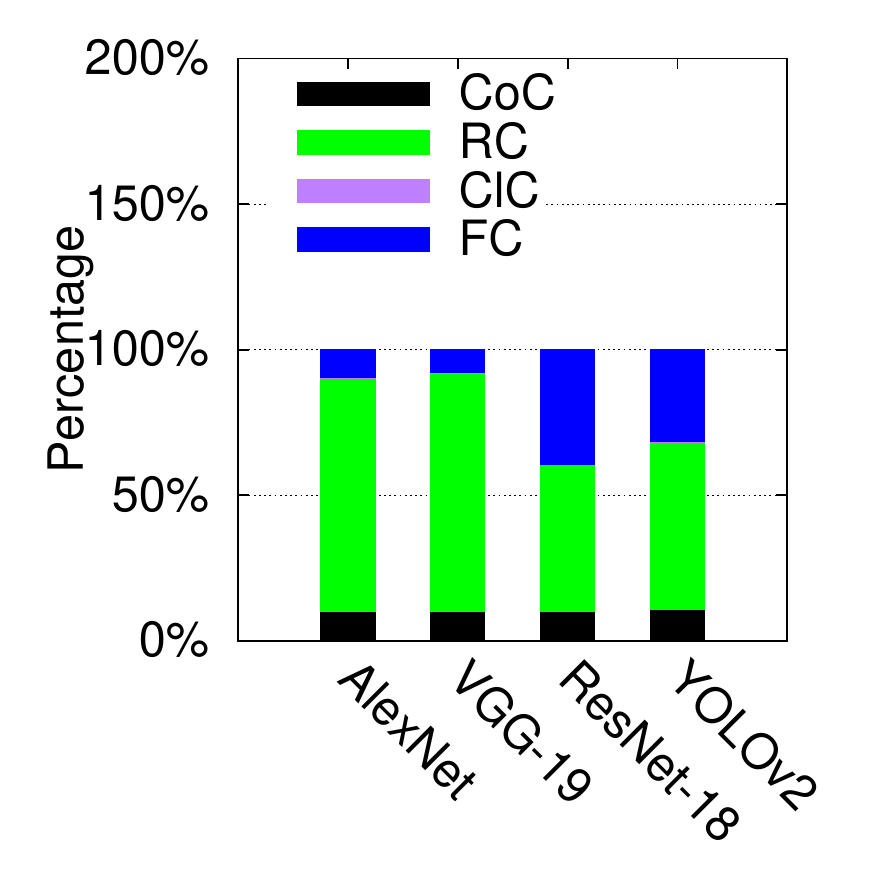}}
}
\hspace{-6mm}
\vspace{-4mm}
\caption{Breakdown Analysis of Multischeme Workflow with MKL-DNN}
\label{fig:statistics_mkldnn}
\end{figure}

\emph{Erroneous cases --  layerwise RC/ClC optimization.} 
Figure \ref{fig:overhead_mkldnn}(c) demonstrates the runtime overhead with layerwise optimization enabled. Every layer decides whether to use RC/ClC independently, as described in Section \ref{sec: solution-sub}. Compared with Figure \ref{fig:overhead_mkldnn}(b), the error correction overhead decreases by 40\%$\sim$60\% (e.g., 1.55\% $\rightarrow$ 0.72\% for YOLOv2 as shown in Figure \ref{fig:overhead_mkldnn}(b) vs. (c)) in all CNN models because of the effectiveness of RC. 
Figure \ref{fig:statistics_mkldnn}(a) shows the distribution of varies workflows that is the result of layerwise RC/ClC optimization. We can see that RC is enabled in all layers of AlexNet and VGG-19, while it is disabled in 30\% to 40\% of layers in ResNet-18 and YOLOv2. The results demonstrate the need for layerwise RC optimization since RC is not suitable for all layers in the same CNN model. Figure \ref{fig:statistics_mkldnn}(b) shows the distribution of soft errors by the schemes that correct them. Less than 5\% of soft errors are corrected by CoC because of the low correction ability of CoC. RC corrects nearly 90\% of the soft errors in AlexNet and VGG-19 because RC is enabled in all layers of the two CNN models and the probability of soft errors striking a row in \textbf{O} is higher than the probability of soft errors striking a column.

\begin{figure}[ht] \centering
\hspace{-8mm}
\subfigure[{AlexNet}]
{
\raisebox{-1cm}{\includegraphics[scale=0.37]{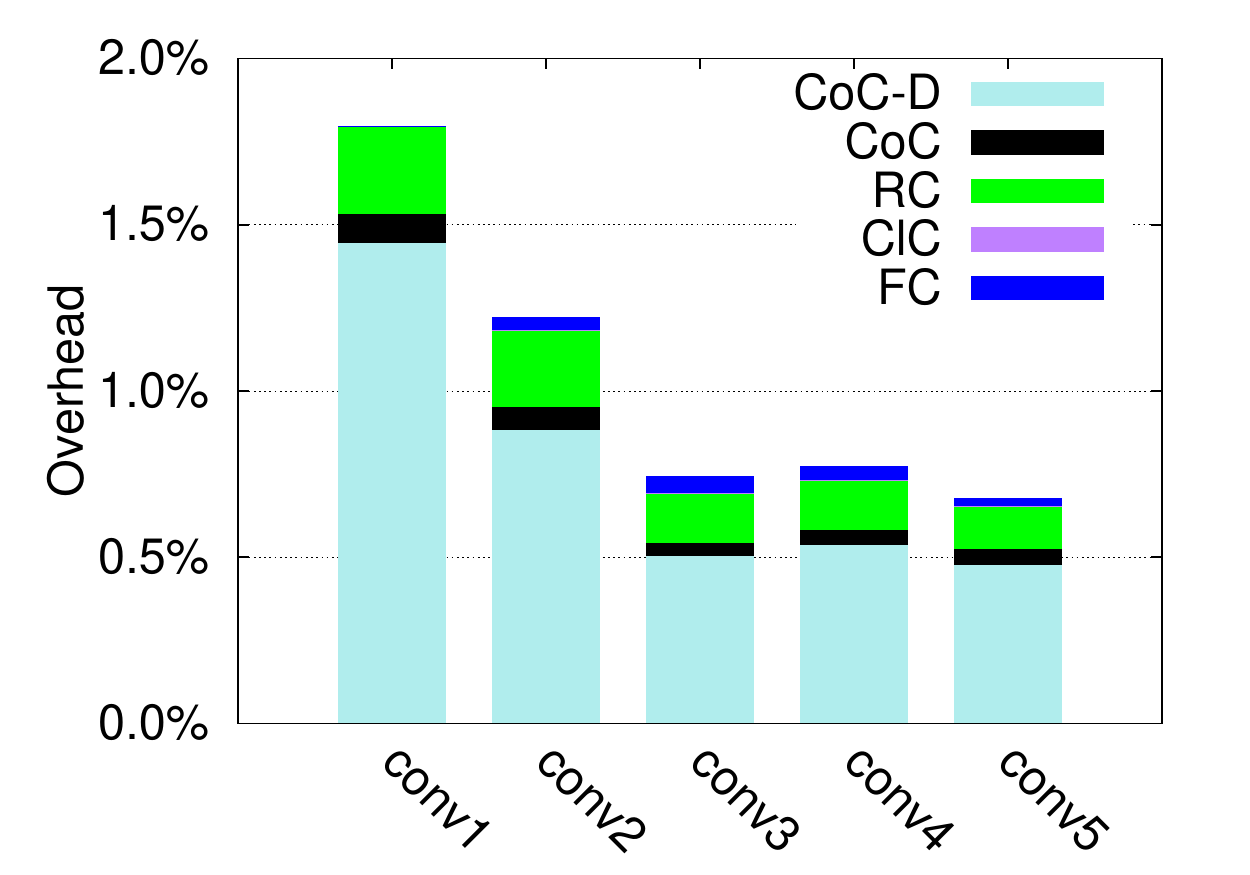}}
}
\hspace{-8mm}
\subfigure[YOLOv2]
{
\raisebox{-1cm}{\includegraphics[scale=0.37]{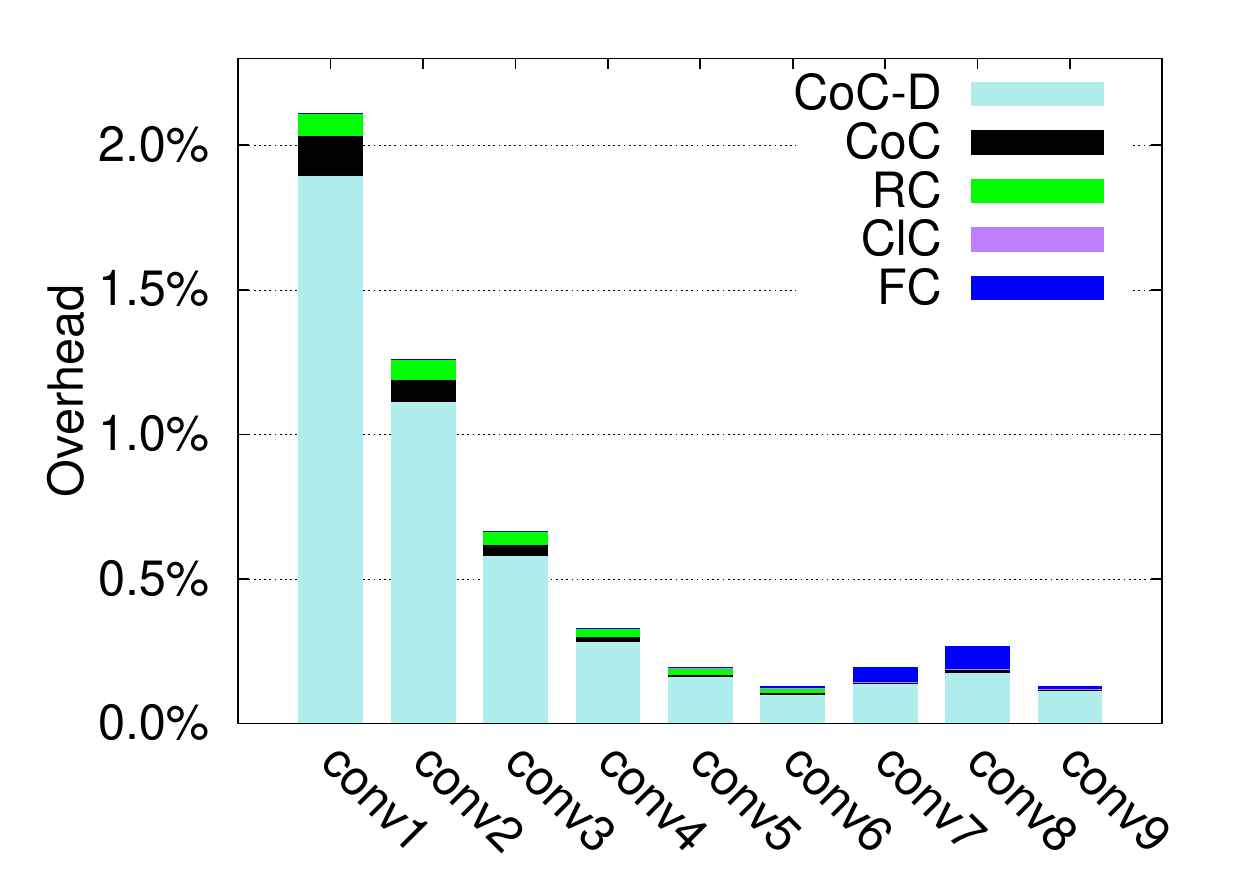}}
}
\hspace{-6mm}
\vspace{-3mm}

\hspace{-8mm}
\subfigure[{VGG-19}]
{
\raisebox{-1cm}{\includegraphics[scale=0.37]{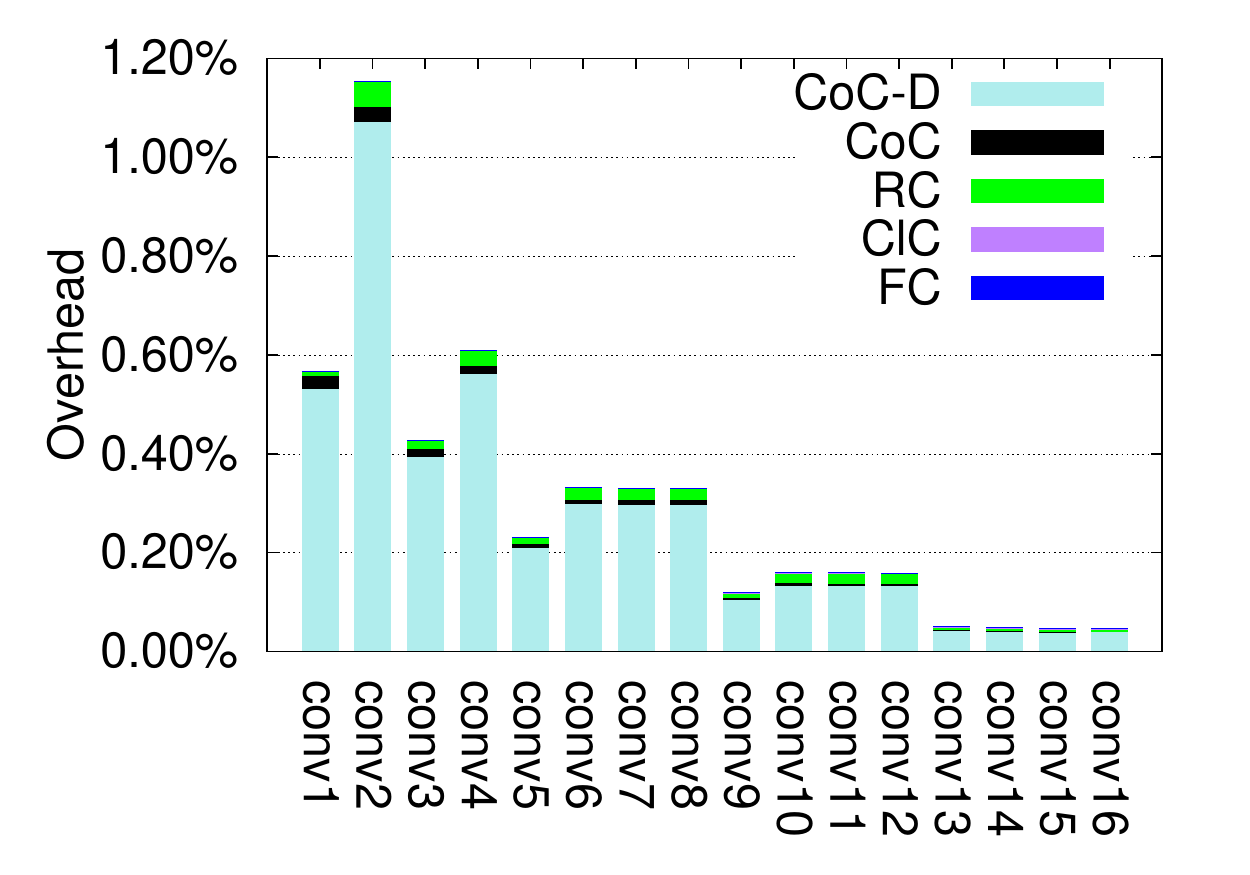}}
}
\hspace{-8mm}
\subfigure[ResNet-18]
{
\raisebox{-1cm}{\includegraphics[scale=0.37]{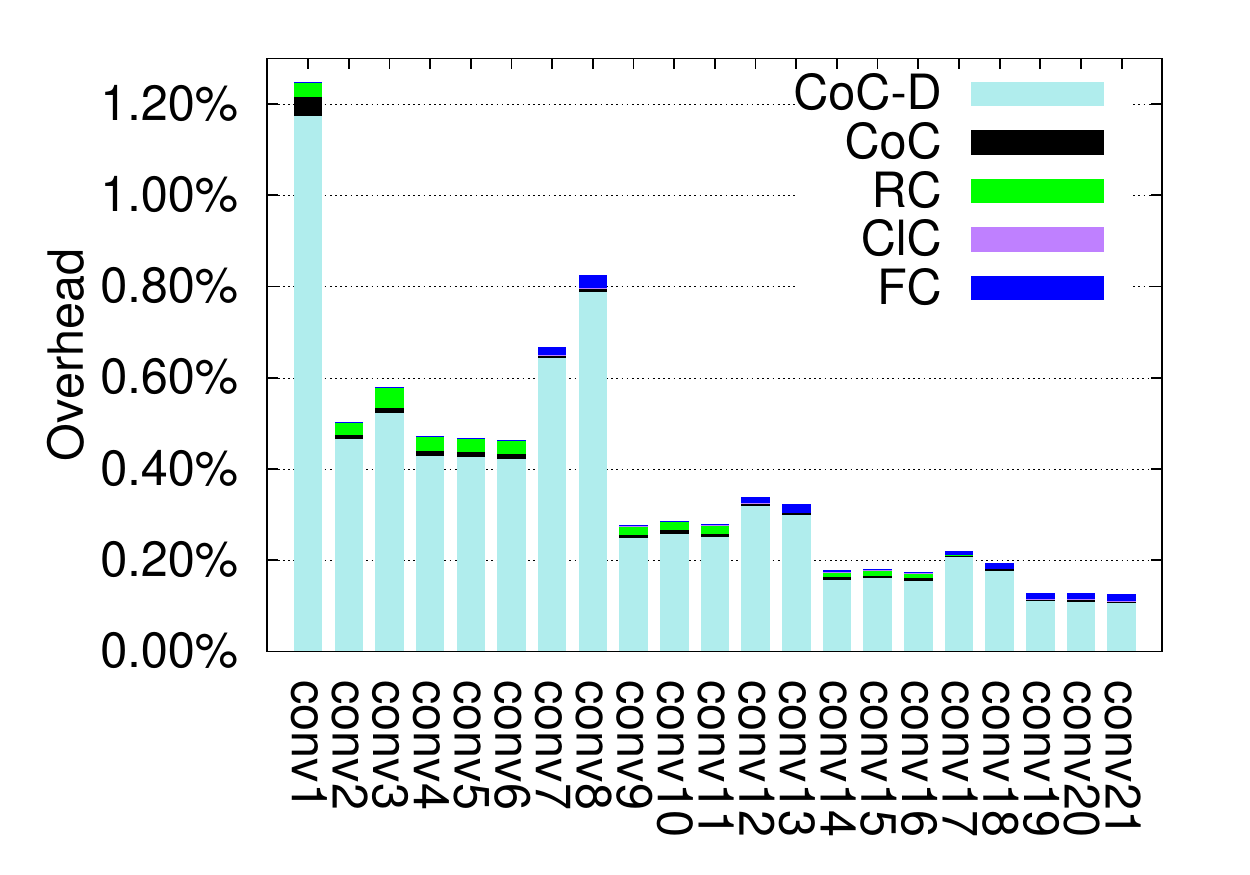}}
}
\hspace{-6mm}
\vspace{-3mm}
\caption{Breakdown of Runtime Overhead by Layer with MKL-DNN}
\label{fig:layerwise_overhead}
\end{figure}

\emph{Erroneous cases -- breakdown of error correction overhead by layer.} To better illustrate the overhead of our solution for each model, we present in Figure \ref{fig:layerwise_overhead} the breakdown of the overhead by layer. The figure demonstrates that layers have diverse overheads of error protection due to the different shapes of \textbf{D} and \textbf{W}. We also notice that the overhead of RC differs among layers in the same model, thus confirming the functionality of our layerwise RC/ClC optimization.

\subsection{Experimental Results with MM-Based Convolution}
\label{sec:mmft_overhead} 

In this section, we evaluate the runtime overhead of our multischeme workflow and the traditional MM-based ABFT. 
Since the MM-based ABFT supports only the MM-based convolution implementation, we set the convolution implementation to the MM-based mode in MKL-DNN. We implemented MM-based ABFT rigorously based on \cite{wu2016}, 
which has $\leq$1\% overhead for large and square matrices as claimed by the authors of that work.
The overhead of the MM-based ABFT in convolution execution is shown in Table \ref{tab:overhead_mmft}. The MM-based ABFT incurs up to 60\% overhead even without error injection for the four CNN models. This result is consistent with our analysis in Section \ref{sec: abft_mm_intro}. Considering that the MM-based ABFT cannot protect the whole process of MM-based convolution and cannot protect other convolution implementations, we  conclude that the MM-based ABFT is unsuitable for soft error protection of CNN applications. 

\begin{figure}[ht] \centering
\hspace{-10mm}
\subfigure[{Error-Free}]
{
\raisebox{-1cm}{\includegraphics[scale=0.4]{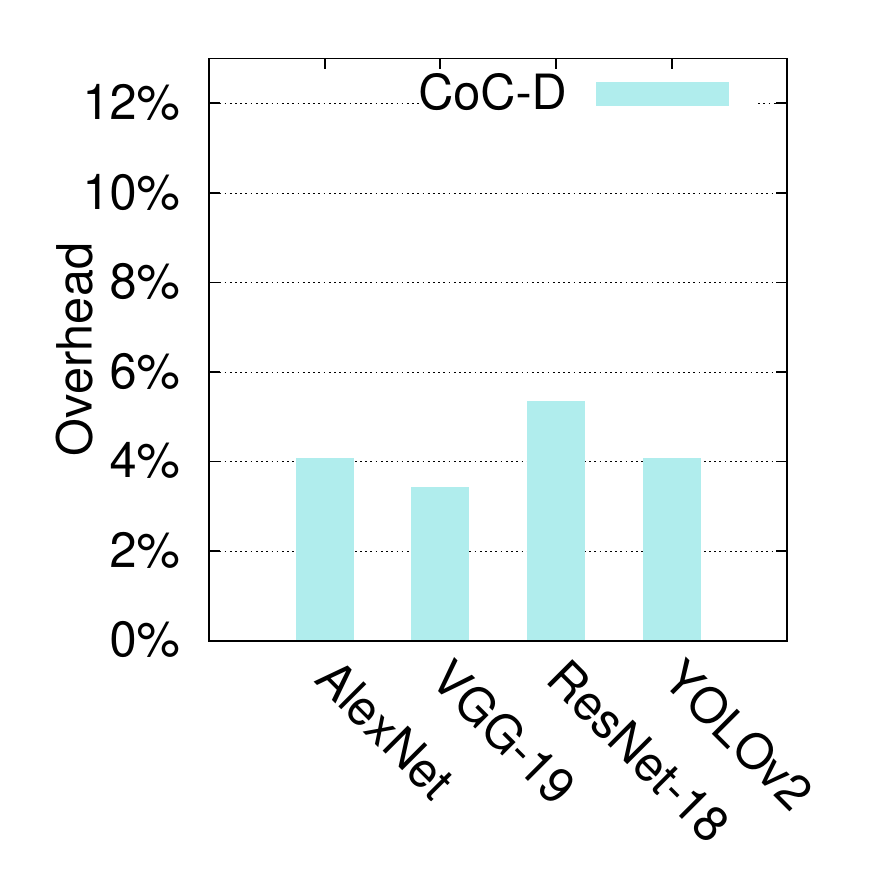}}
}
\hspace{-9mm}
\subfigure[{\hspace*{8mm} Erroneous \newline \hspace*{8mm}(RC/ClC Disabled)}]
{
\raisebox{-1cm}{\includegraphics[scale=0.4]{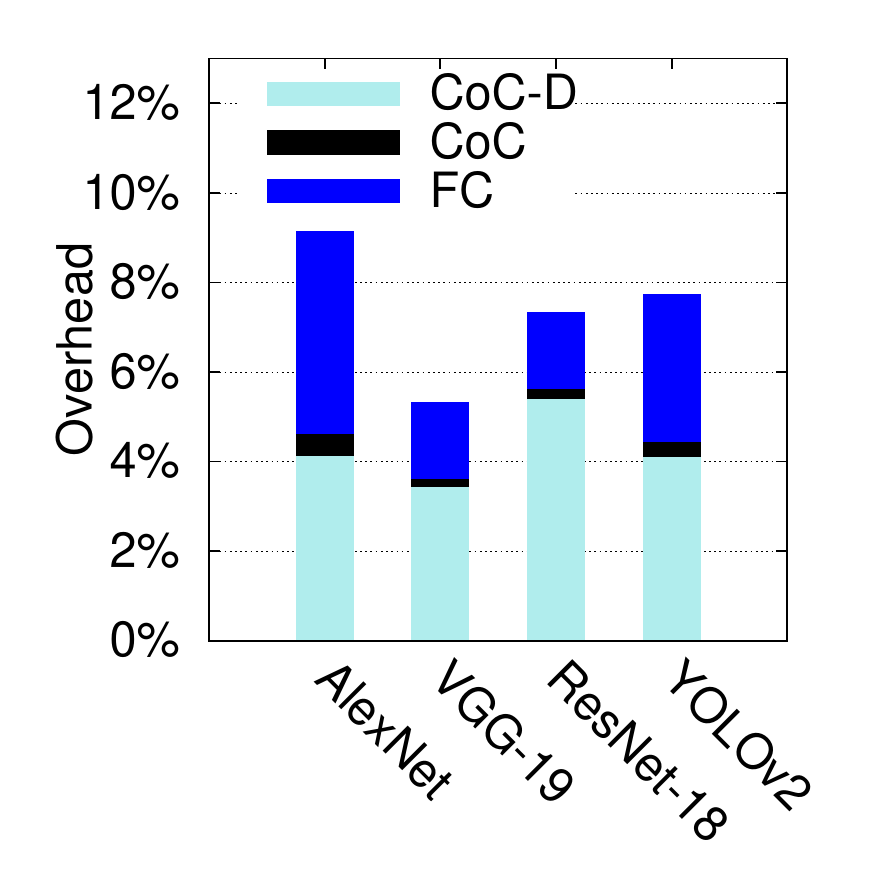}}
}
\hspace{-9mm}
\subfigure[{\hspace{8mm} Erroneous \newline \hspace*{8mm}(RC/ClC enabled)}]
{
\raisebox{-1cm}{\includegraphics[scale=0.4]{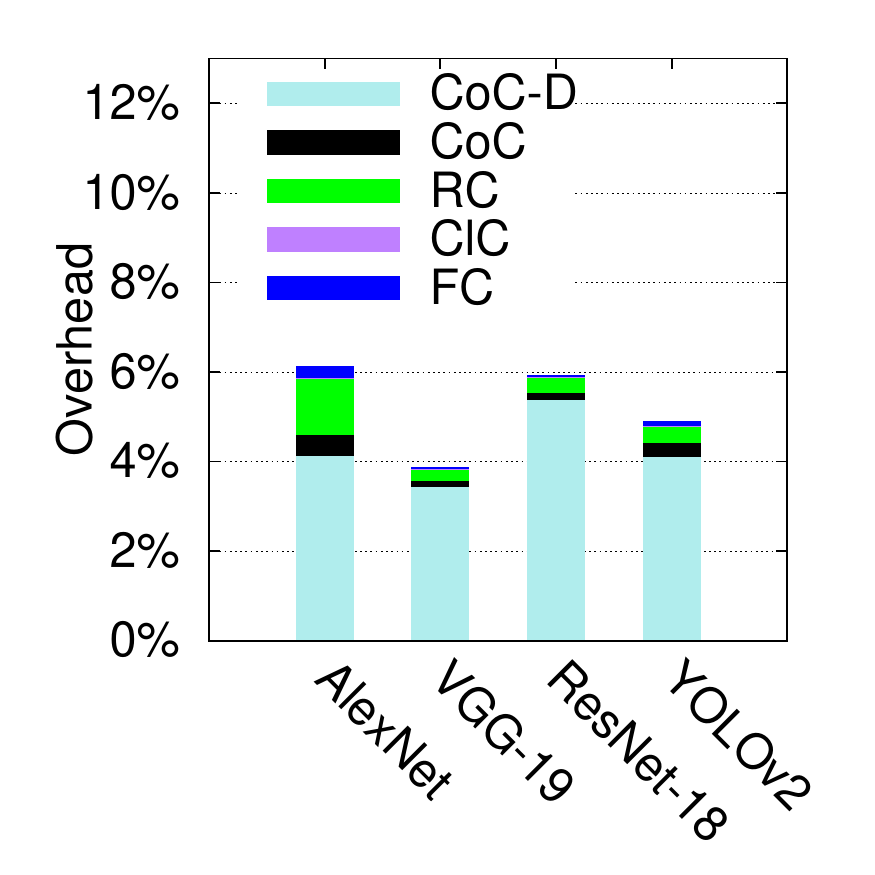}}
}
\hspace{-8mm}
\vspace{-3mm}
\caption{Runtime Overhead with MM-based Convolution}
\label{fig:overhead_gemm}
\end{figure}

\begin{figure}[ht] \centering
\vspace{-5mm}
\hspace{-3mm}
\subfigure[Distribution of different workflows]
{
\raisebox{-1cm}{\includegraphics[scale=0.45]{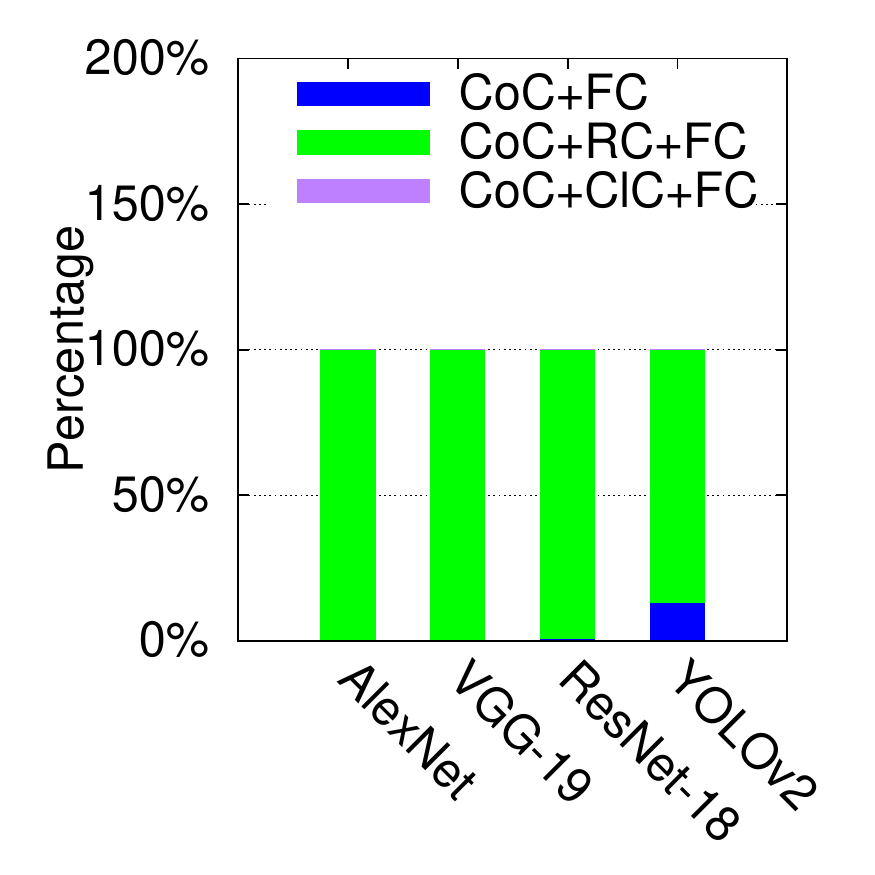}}
}
\hspace{4mm}
\subfigure[{Distribution of soft errors corrected by schemes}]
{
\raisebox{-1cm}{\includegraphics[scale=0.45]{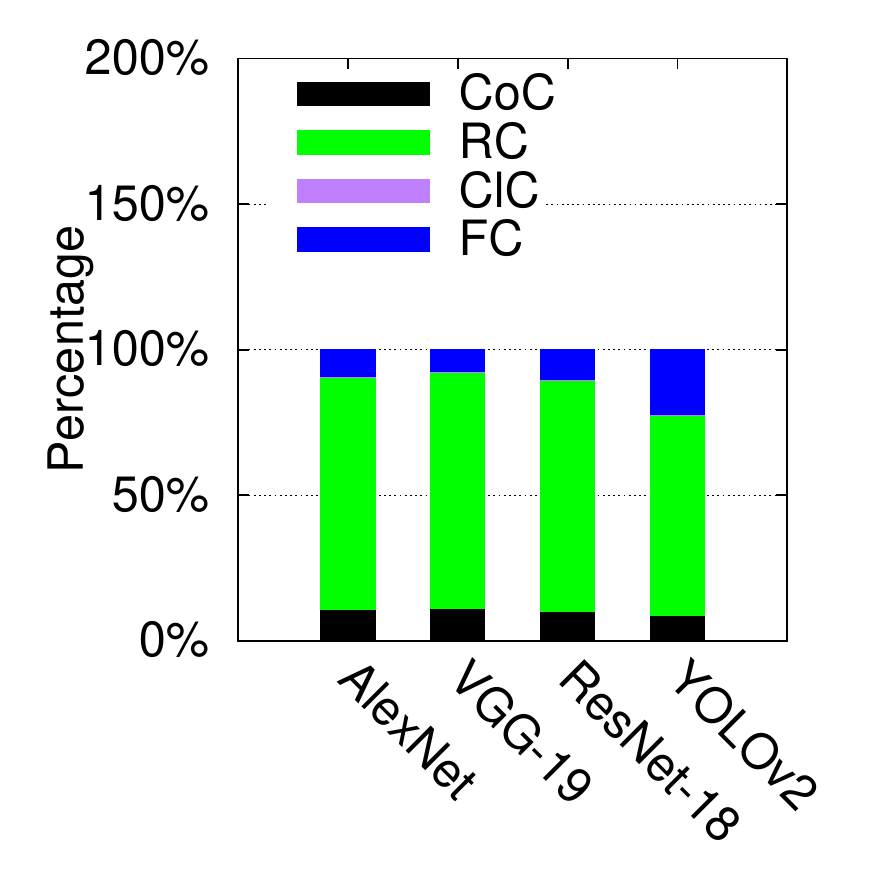}}
}
\hspace{-4mm}
\vspace{-3mm}
\caption{Breakdown Analysis of Multischeme Workflow with MM-Based Convolution}
\label{fig:statistics_gemm}
\end{figure}

Figure \ref{fig:overhead_gemm} shows the overhead of our multischeme workflow for MM-based convolution. The overhead of our solution is below 6\% in the error-free cases and below 6.5\% in the cases with injected errors for all CNN models. The layerwise RC/ClC optimization reduces the overhead for error correction by as much as 77\%. Figure \ref{fig:statistics_gemm}(a) shows the fractions of different workflows chosen by convolutional layers. Figure \ref{fig:statistics_gemm}(b) shows the distribution of soft errors that are corrected by different schemes. Compared with the MKL-DNN implementation, more layers adopt RC for error correction in the MM-based convolution (see  Figure \ref{fig:statistics_mkldnn} versus  Figure \ref{fig:statistics_gemm}). The reason is that the relative runtime of RC compared with FC is lower in the MM-based convolution implementation than other implementations. 

\begin{table}[ht]
\centering
\footnotesize
  \caption {Overhead of MM-Based ABFT for MM-Based Convolution, No Error Injection} 
  \vspace{-4mm}
  \label{tab:overhead_mmft} 
  \begin{adjustbox}{width=0.9\columnwidth}
\begin{tabular}{|c|c|c|c|c|}
\hline
Model    & AlexNet & YOLOv2 & VGG-19 & ResNet-18 \\ \hline
Overhead & 27.9\%  & 57.5\% &  45.8\%      &        61.2\%   \\ \hline
\end{tabular}
\end{adjustbox}
\end{table}

\subsection{Parallel Performance Evaluation}
In this section, we present the parallel performance evaluation results of AlexNet, YOLOv2, VGG-19, and ResNet-18. 
Original images of the ImageNet validation dataset are used without preprocessing in order to better demonstrate the process of parallel CNN inference application. In the beginning of the parallel process, images are distributed to the local disk of each node; then each node starts to do the data processing step first to convert the images to suitable size required by CNN models, and then execute the inference step under the protection of our multischeme workflow.

\begin{figure}[ht] \centering
\hspace{-8mm}
\subfigure[{AlexNet}]
{
\raisebox{-1cm}{\includegraphics[scale=0.37]{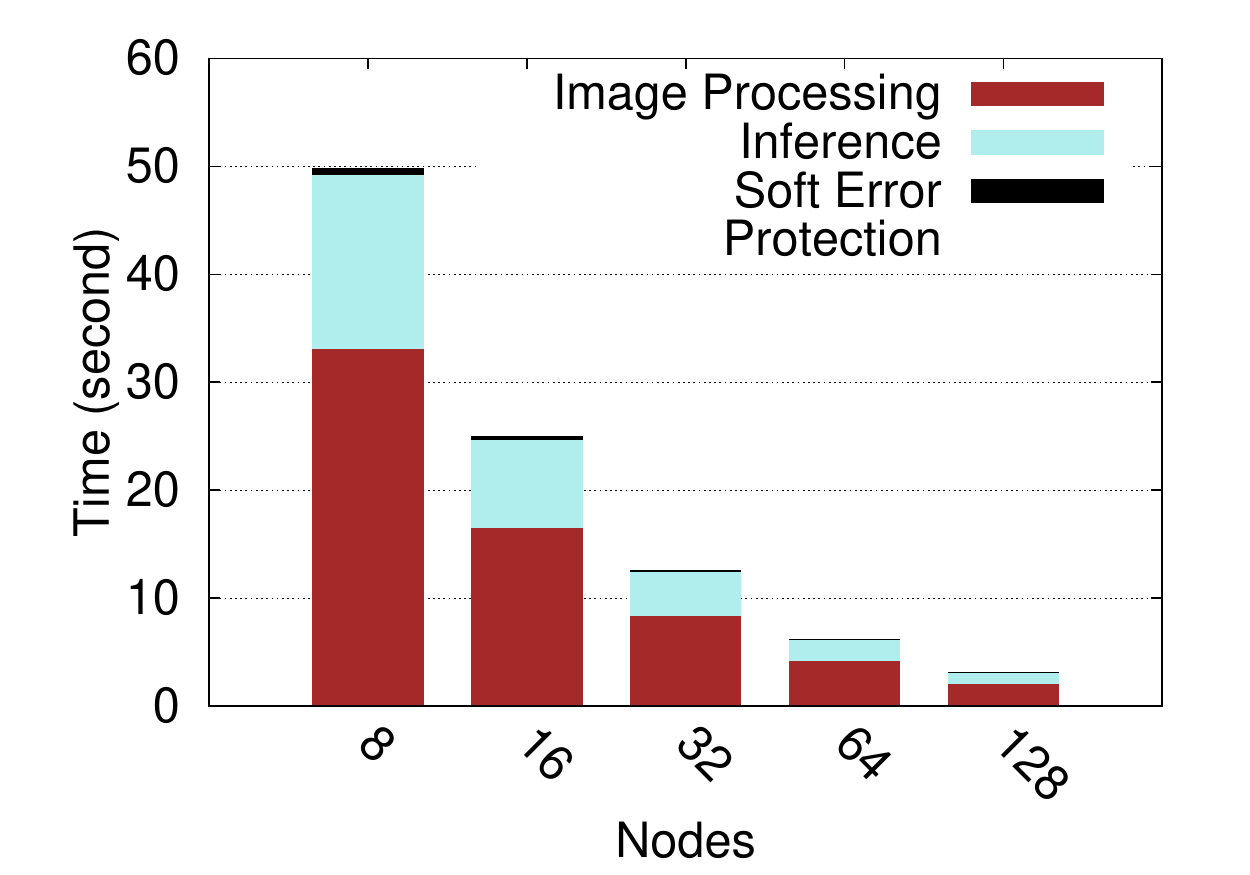}}
}
\hspace{-8mm}
\subfigure[YOLOv2]
{
\raisebox{-1cm}{\includegraphics[scale=0.37]{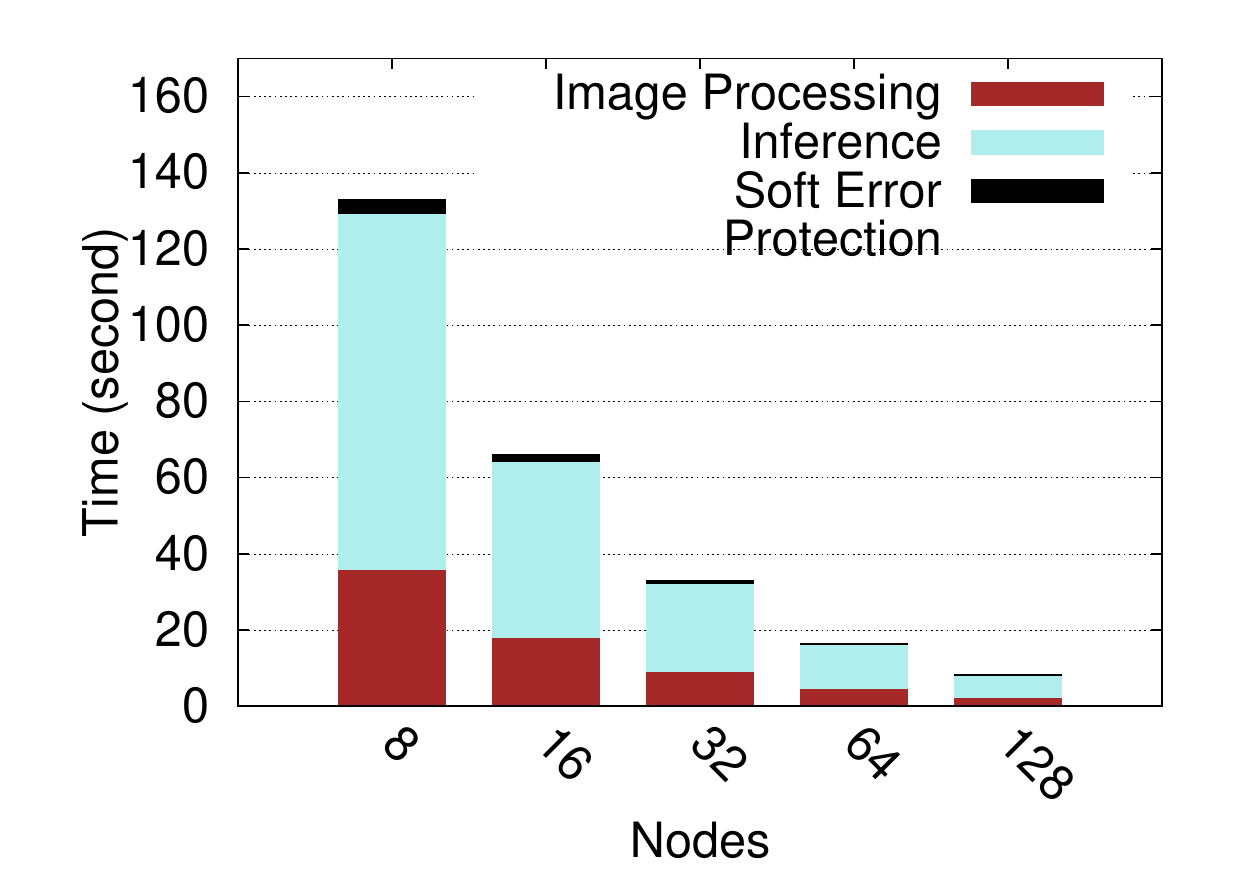}}
}
\hspace{-6mm}
\vspace{-1mm}

\hspace{-8mm}
\subfigure[{VGG-19}]
{
\raisebox{-1cm}{\includegraphics[scale=0.37]{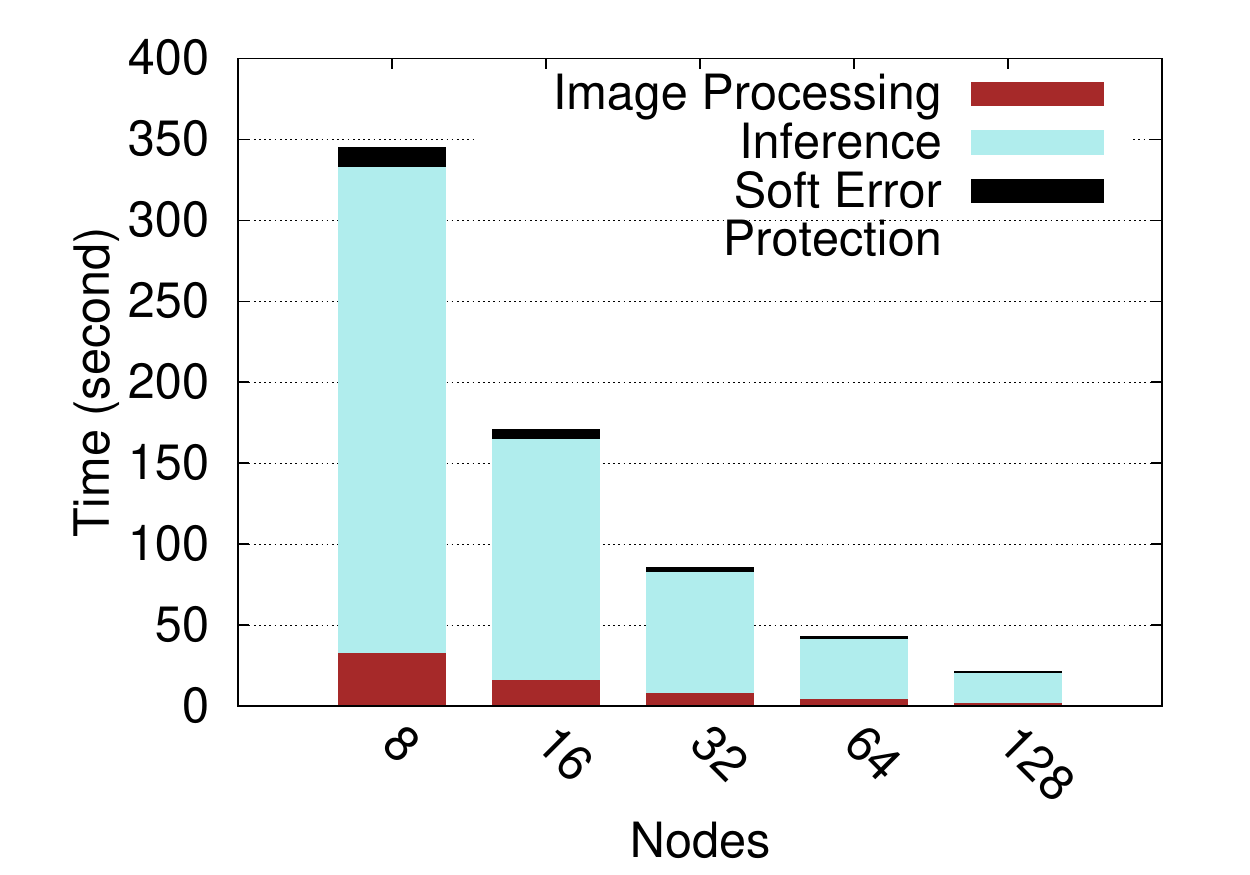}}
}
\hspace{-8mm}
\subfigure[{ResNet-18}]
{
\raisebox{-1cm}{\includegraphics[scale=0.37]{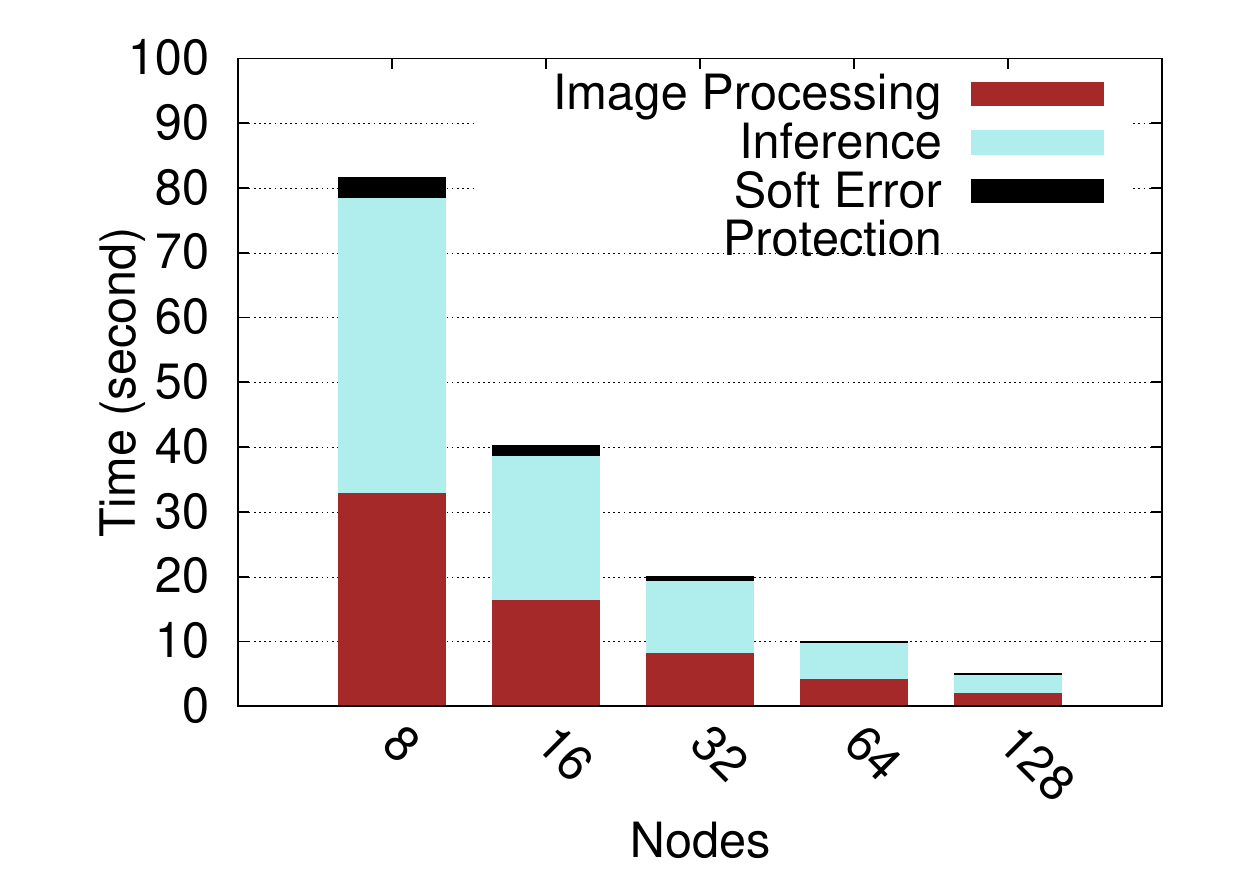}}
}
\hspace{-6mm}
\vspace{-2mm}
\caption{Parallel Performance Evaluation of Our Solution with Injected Errors on Bebop Supercomputer}
\label{fig:parallel_eva}
\end{figure}

We conducted the parallel evaluation in both error-free and erroneous cases. However, because of space limits, we present only the parallel performance evaluation results in the situation with injected errors (as shown in Figure \ref{fig:parallel_eva}). In fact, the evaluation results in the error-free situation are  similar. Specifically, experiments show that our multischeme workflow has a very good scalability:  that is, the soft error protection overhead does not increase with the number of nodes at all. In absolute terms, the overhead stays around 2\%$\sim$6\% in the erroneous cases and is only 1\%$\sim$4\% in the error-free cases.

\vspace{-3mm}

\section{Related Work}
\label{sec: related work}
The importance of fault tolerance for convolution has been emerging in recent years. Guaranteeing the correctness of inference is  vital in a safety-critical use case \cite{Li_SC17}. To achieve  better resiliency for CNN networks, researchers have been exploring solutions from different perspectives including hardware, system, and software. For hardware, Kim et al. \cite{Kim:2019:DDN:3316781.3317805} proposed a hardened 3D die-stacked memory based on the fault characteristics in convolutional DNNs. Li et al. \cite{Li_SC17} proposed to add redundant circuits selectively to harden the latches based on analysis of data resiliency. Compared with traditional full-hardware redundancy techniques, those partial-hardware redundancy techniques may not double the power usage. However, hardware modification incurs significant effort considering the varied CNN models and their accelerators. At the system level, other than the DMR/TMR protection, checkpoint/restart (C/R) is also applied to  large-scale machine learning systems. Subsequently,  Qiao et al. proposed a more efficient C/R scheme based on their derived upper bound on extra iteration cost with perturbations \cite{DBLP:journals/corr/abs-1810-07354}. While those C/R techniques are promising to protect model training from soft errors, they are not good fits for inference since one inference execution could be very fast and applying C/R incurs significant overhead. Researchers have therefore pursued lightweight software-level solutions. By applying ABFT techniques for MM-based convolution, Santos et al. \cite{8023727} reported that 50\%$\sim$60\% of radiation-induced corruptions could be corrected. Unfortunately, the traditional ABFT works only for MM-based convolution, which is inefficient in most cases. 
In contrast, our solutions can work for any convolution implementations.
\vspace{-3mm}
\section{Conclusion and Future Work}
\label{sec: conclusion}
This work focus on extending ABFT to convolution operations in convolutional neural networks. We propose four ABFT schemes and a multischeme workflow to protect the convolutional layer. We further extend our schemes to support bias, grouped convolution, and convolution back propagation. We implement an efficient CNN framework, FT-Caffe, that is resilient to silent data corruption. 

Experiments demonstrate that our proposed fault-tolerant solutions incur negligible overhead. In absolute terms, FT-Caffe can acheive less than 8\% overhead for the most widely used CNN models, including AlexNet, YOLO, VGG-19, and ResNet-18, in both error-free and erroneous cases. 

We plan to extend the implementation to more CNN frameworks and to design architecture-specific optimizations for different hardware including GPU, FPGA, and AI accelerators.


\section*{ACKNOWLEDGMENTS}
This research was supported by the Exascale Computing Project (ECP), Project Number: 17-SC-20-SC, a collaborative effort of two DOE organizations - the Office of Science and the National Nuclear Security Administration, responsible for the planning and preparation of a capable exascale ecosystem, including software, applications, hardware, advanced system engineering and early testbed platforms, to support the nation’s exascale computing imperative. The material was supported by the U.S. Department of Energy, Office of Science, under contract DE-AC02-06CH11357. This work was also supported by the National Science Foundation under Grants CCF-1513201, CCF-1619253, and OAC-2034169. We acknowledge the computing resources provided on Bebop, which is operated by the Laboratory Computing Resource Center at Argonne National Laboratory.

\ifCLASSOPTIONcaptionsoff
  \newpage
\fi



%
\bibliographystyle{IEEEtran}
\bibliography{bib/refs}


%

\begin{IEEEbiography}[{\includegraphics[width=1in,height=1.25in,clip,keepaspectratio]{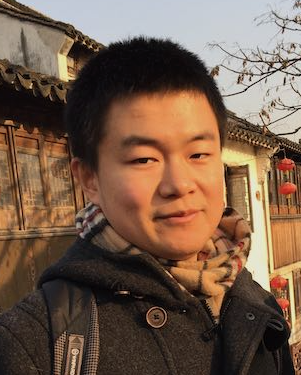}}]{Kai Zhao}
received his bachelor's degree from Peking University in 2014 and will receive his Ph.D. degree from University of California, Riverside in 2022. He is a long-term intern at Argonne National Laboratory.
His research interests include high-performance computing, scientific data management and reduction, and resilient machine learning. Email: kzhao016@ucr.edu.
\end{IEEEbiography}

\vskip -2\baselineskip plus -1fil

\begin{IEEEbiography}[{\includegraphics[width=1in,height=1.25in,clip,keepaspectratio]{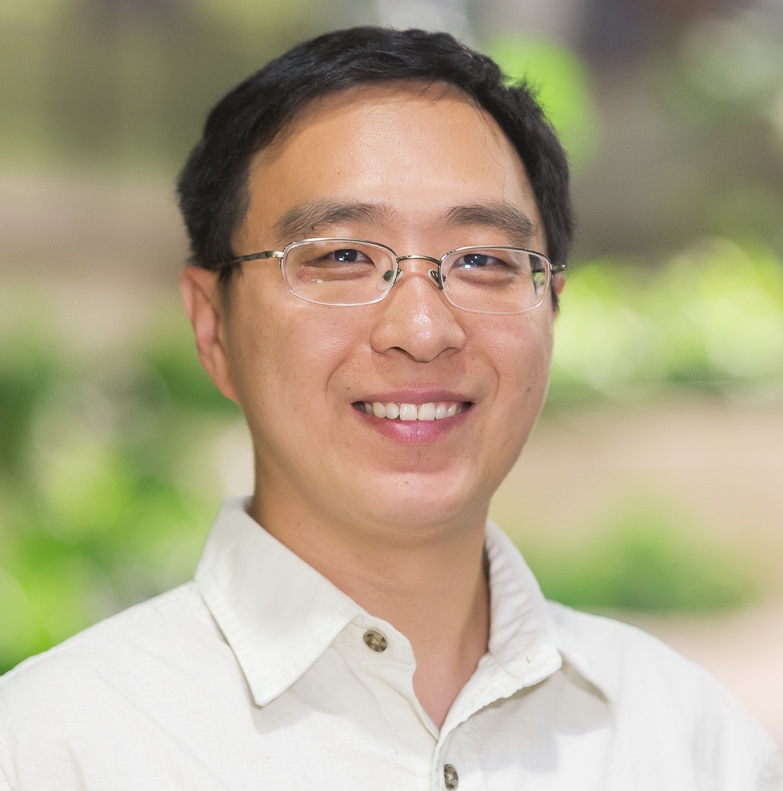}}]{Sheng Di}
(Senior Member, IEEE) received his master's degree from Huazhong University of Science and Technology in 2007 and Ph.D. degree from the University of Hong Kong in 2011. He is currently a computer scientist at Argonne National Laboratory. 
Dr. Di's research interest involves resilience on high-performance computing (such as silent data corruption, optimization checkpoint model, and in-situ data compression) and broad research topics on cloud computing (including optimization of resource allocation, cloud network topology, and prediction of cloud workload/hostload).
He is working on multiple HPC projects, such as detection of silent data corruption, characterization of failures and faults for HPC systems, and optimization of multilevel checkpoint models. Email: sdi1@anl.gov.
\end{IEEEbiography}

\vskip -2\baselineskip plus -1fil

\begin{IEEEbiography}[{\includegraphics[width=1in,height=1.25in,clip,keepaspectratio]{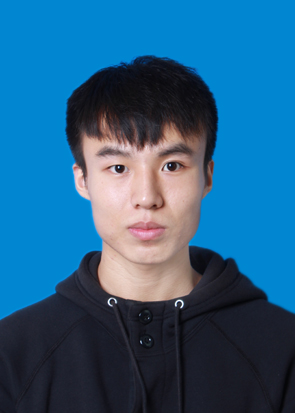}}]{Sihuan Li}
is a Ph.D. student in computer science at University of California, Riverside. He obtained his bachelor's degree in math from Huazhong University of Science and Technology, China. He did a long-term internship at Argonne National Laboratory. Broadly speaking, his research interests fall into High Performance Computing. Specifically, he mainly studies Algorithm Based Fault Tolerance (ABFT), lossy compression and their applications in large scale scientific simulations. He is an IEEE student member. Email: sli049@ucr.edu.
\end{IEEEbiography}

\vskip -2\baselineskip plus -1fil

\begin{IEEEbiography}[{\includegraphics[width=1in,height=1.25in,clip,keepaspectratio]{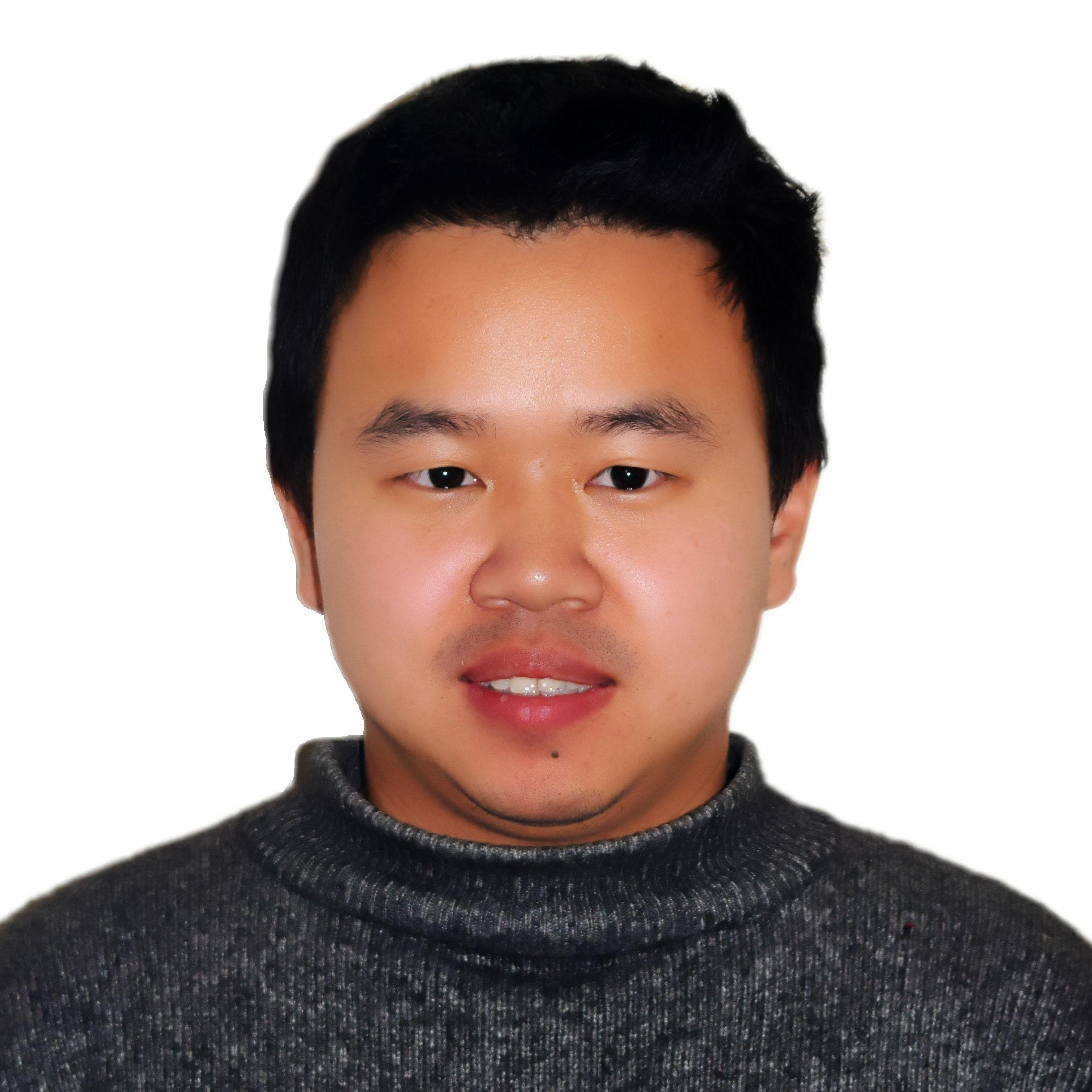}}]{Xin Liang} is a Computer/Data Scientist at Oak Ridge National Laboratory. He received his Ph.D. degree from University of California, Riverside in 2019 and his bachelor's degree from Peking University in 2014. His research interests include high-performance computing, parallel and distributed systems, scientific data management and reduction, big data analytic, scientific visualization, and cloud computing. He has interned in multiple national laboratories and worked on several exascale computing projects. He is a member of the IEEE. Email: liangx@ornl.gov.
\end{IEEEbiography}

\vskip -2\baselineskip plus -1fil

\begin{IEEEbiography}[{\includegraphics[width=1in,height=1.25in,clip,keepaspectratio]{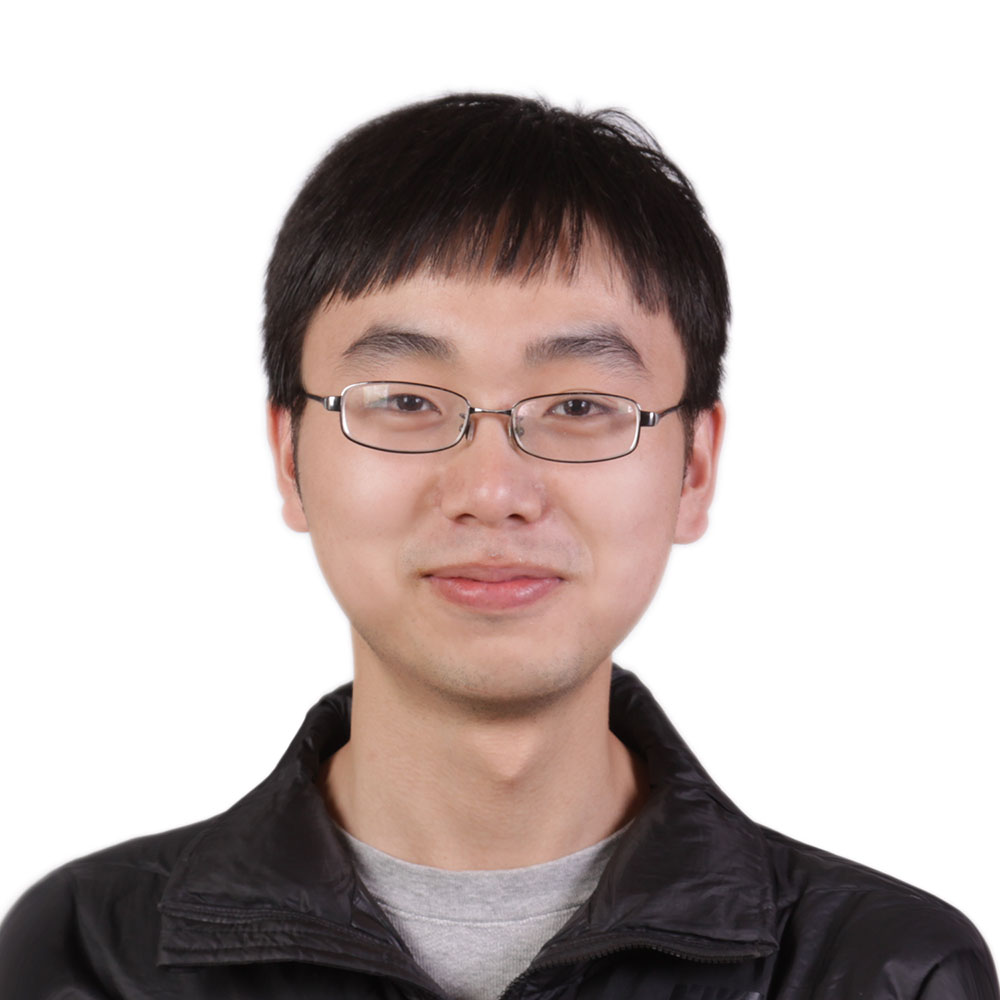}}]{Yujia Zhai}
received his bachelor's degree from University of Science and Technology of China in 2016, a master's degree from Duke University in 2018, and will receive his Ph.D. degree from University of California, Riverside in 2023. His research interests include high-performance computing, parallel and distributed systems, and numerical linear algebra software. Email: yzhai015@ucr.edu.
\end{IEEEbiography}

\vskip -2\baselineskip plus -1fil

\begin{IEEEbiography}[{\includegraphics[width=1in,height=1.25in,clip,keepaspectratio]{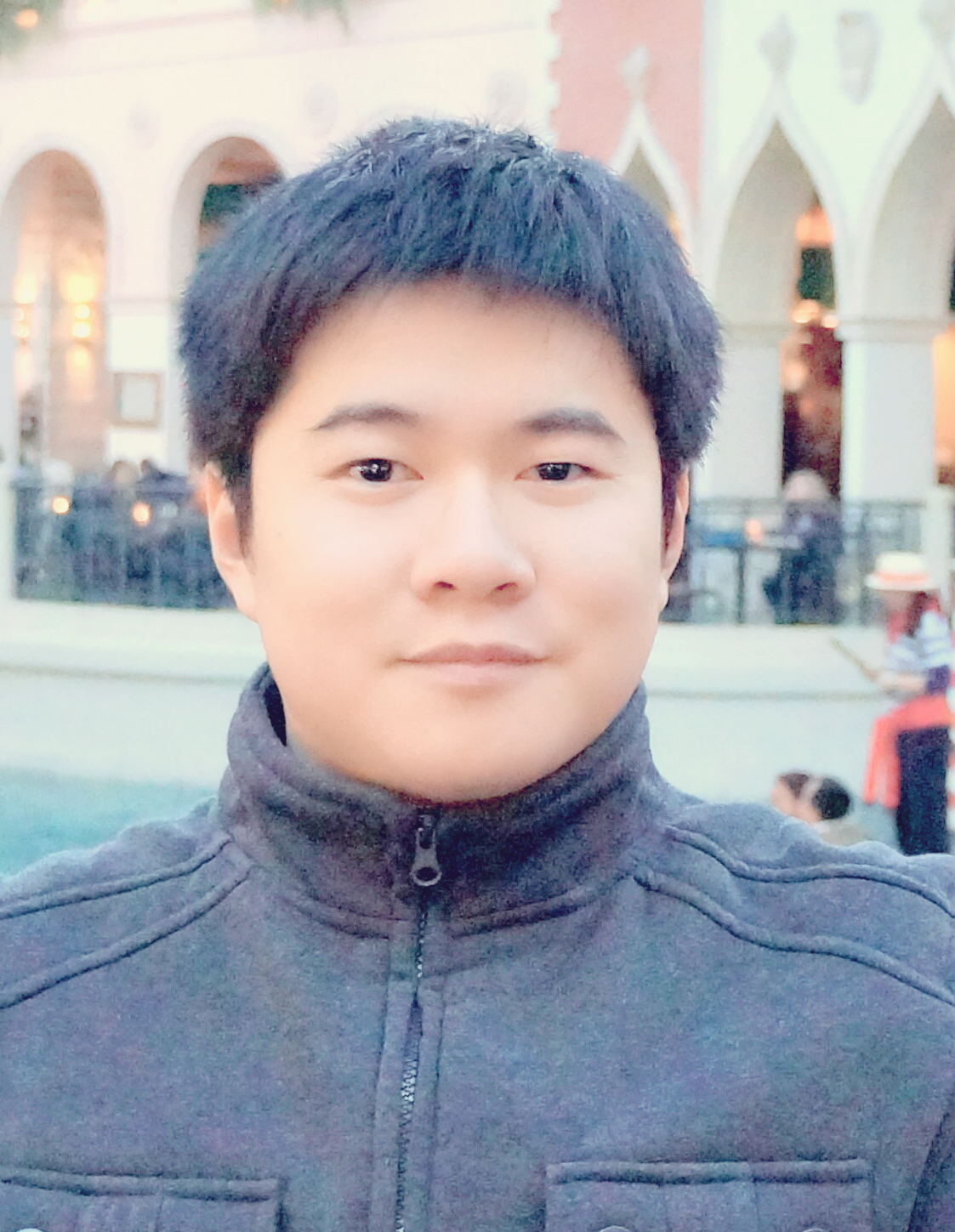}}]{Jieyang Chen} (Member, IEEE) is a Computer Scientist at Computer Science and Mathematics Division at Oak Ridge National Laboratory.
He received his master and Ph.D. degrees in Computer Science from University of California, Riverside in 2014 and 2019. He received a bachelor’s degree in Computer Science and Engineering from Beijing University of Technology in 2012. His research interests include high-performance computing, parallel and distributed systems, and big data analytics. Email: chenj3@ornl.gov.
\end{IEEEbiography}

\vskip -2\baselineskip plus -1fil

\begin{IEEEbiography}[{\includegraphics[width=1in,height=1.25in,clip,keepaspectratio]{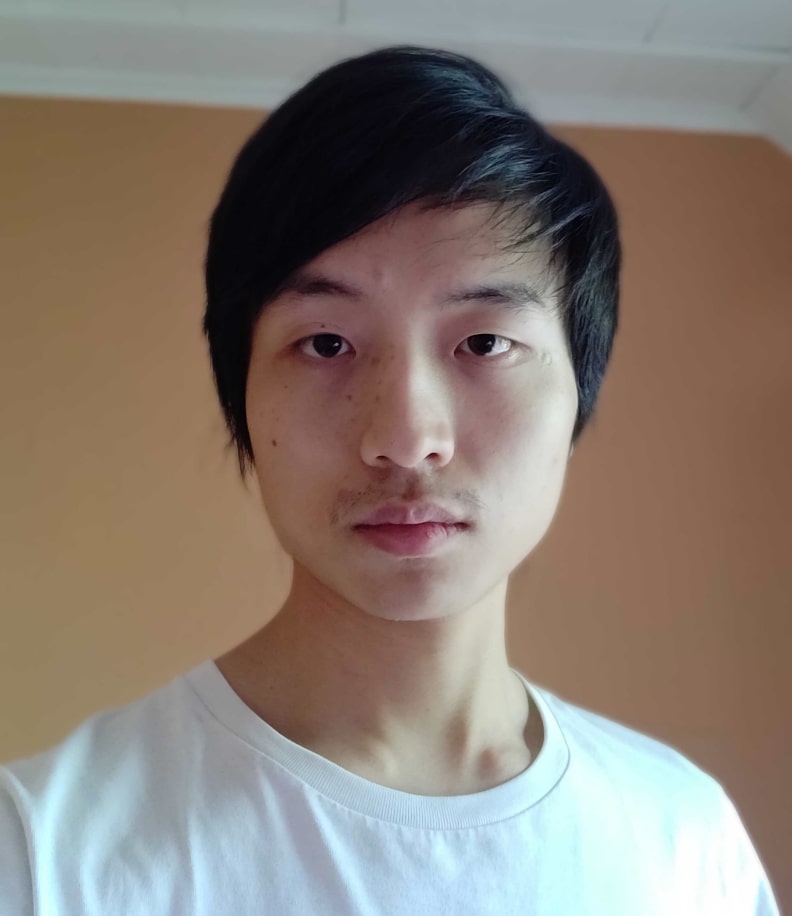}}]{Kaiming Ouyang}
received his bachelor's degree from University of Electronic Science and Technology of China and joined the University of California, Riverside SuperLab in Fall 2016. 
He will receive his Ph.D. degree from University of California, Riverside in 2021. He is a long-term intern at Argonne National Laboratory PMRS group led by Dr. Balaji and supervised by Dr. Si. His research interest is parallel runtime system. Email: kouya001@ucr.edu.
 \end{IEEEbiography}

\vskip -2\baselineskip plus -1fil

\begin{IEEEbiography}[{\includegraphics[width=1in,height=1.25in,clip,keepaspectratio]{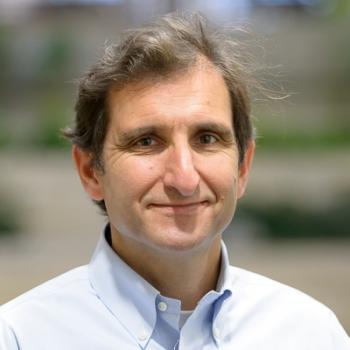}}]{Franck Cappello}
(Fellow, IEEE) is the director of the Joint-Laboratory on Extreme Scale Computing gathering six of the leading high-performance computing institutions in the world: Argonne National Laboratory, National Center for Scientific Applications, Inria, Barcelona Supercomputing Center, Julich Supercomputing Center, and Riken AICS. He is a senior computer scientist at Argonne National Laboratory and an adjunct associate professor in the Department of Computer Science at the University of Illinois at Urbana-Champaign. He is an expert in resilience and fault tolerance for scientific computing and data analytics. Recently he started investigating lossy compression for scientific data sets to respond to the pressing needs of scientist performing large-scale simulations and experiments. His contribution to this domain is one of the best lossy compressors for scientific data set respecting user-set error bounds. He is a member of the editorial board of the \textit{IEEE Transactions on Parallel and Distributed Computing} and of the \textit{ACM HPDC} and \textit{IEEE CCGRID} steering committees. He is a fellow of the IEEE. Email: cappello@mcs.anl.gov.
\end{IEEEbiography}

\vskip -2\baselineskip plus -1fil

\begin{IEEEbiography}[{\includegraphics[width=1in,height=1.25in,clip,keepaspectratio]{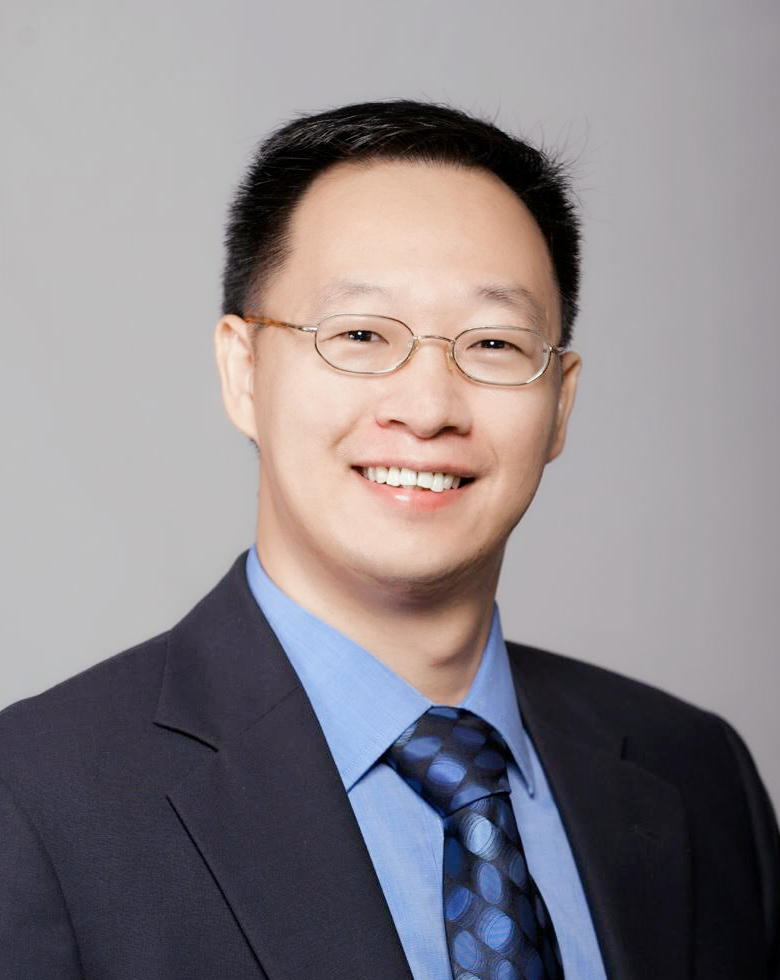}}]{Zizhong Chen}
(Senior Member, IEEE) received a bachelor's degree in mathematics from Beijing Normal University, a master's degree degree in economics from the Renmin University of China, and a Ph.D. degree in computer science from the University of Tennessee, Knoxville. He is a professor of computer science at the University of California, Riverside. 
His research interests include high-performance computing, parallel and distributed systems, big data analytics, cluster and cloud computing, algorithm-based fault tolerance, power and energy efficient computing, numerical algorithms and software, and large-scale computer simulations. His research has been supported by National Science Foundation, Department of Energy, CMG Reservoir Simulation Foundation, Abu Dhabi National Oil Company, Nvidia, and Microsoft Corporation. 
He received a CAREER Award from the US National Science Foundation and a Best Paper Award from the International Supercomputing Conference. 
He is a Senior Member of the IEEE and a Life Member of the ACM. He currently serves as a subject area editor for \textit{Elsevier Parallel Computing} journal and an associate editor for the \textit{IEEE Transactions on Parallel and Distributed Systems}. Email: chen@cs.ucr.edu.
\end{IEEEbiography}






\end{document}